\documentclass[iop]{emulateapj}

\newcommand{\Msolar}{\mbox{\,$\rm M_{\odot}$}}   
\newcommand{\civ}{C\,{\sc iv}} 
\newcommand{\mgii}{Mg\,{\sc ii}}
\newcommand{\feii}{Fe\,{\sc ii}} 
\newcommand{\ciii}{C\,{\sc iii]}} 
\newcommand{\nv}{N\,{\sc v}}
 
\newcommand{\oi}{O\,{\sc i}} 
\newcommand{\siiv}{Si\,{\sc iv}}
\newcommand{\oiv}{O\,{\sc iv]}}
\newcommand{\hi}{H\,{\sc i}}
\usepackage{graphicx}   
\usepackage{amsmath}    
\usepackage{amssymb}    
\usepackage{ulem}      

\shorttitle{IMS III}
\shortauthors{Jeon et al.}

\begin{document}

\title{
The Infrared Medium-deep Survey. III. \\
Survey of Luminous Quasars at 4.7 $\leq$ z $\leq$ 5.4$^{\star}$
}

\author{
Yiseul Jeon\altaffilmark{1,2,$\dagger$,$\ddagger$}, 
Myungshin Im\altaffilmark{1,$\dagger$,$\ddagger$},
Dohyeong Kim\altaffilmark{1,$\dagger$}, 
Yongjung Kim\altaffilmark{1},
Hyunsung David Jun\altaffilmark{1,3},
Soojong Pak\altaffilmark{4}, 
Yoon Chan Taak\altaffilmark{1}, 
Giseon Baek\altaffilmark{4},
Changsu Choi\altaffilmark{1}, 
Nahyun Choi\altaffilmark{5}, 
Jueun Hong\altaffilmark{1}, 
Minhee Hyun\altaffilmark{1}, 
Tae-Geun Ji\altaffilmark{4}, 
Marios Karouzos\altaffilmark{1,6}, 
Duho Kim\altaffilmark{1,7}, 
Jae-Woo Kim\altaffilmark{1,8}, 
Ji Hoon Kim\altaffilmark{1,9},
Minjin Kim\altaffilmark{8}, 
Sanghyuk Kim\altaffilmark{8},
Hye-In Lee\altaffilmark{4},
Seong-Kook Lee\altaffilmark{1}, 
Won-Kee Park\altaffilmark{1,8}, 
Woojin Park\altaffilmark{4},
Yongmin Yoon\altaffilmark{1}
}

\altaffiltext{1}{Center for the Exploration of the Origin of the Universe (CEOU), Astronomy Program, Department of Physics \& Astronomy, Seoul National University, 1 Gwanak-ro, Gwanak-gu, Seoul 151-742 Korea}
\altaffiltext{2}{LOCOOP, Inc., 311-1, 108 Gasandigital2-ro, Geumcheon-gu, Seoul, Korea}
\altaffiltext{3}{Jet Propulsion Laboratory, California Institute of Technology, 4800 Oak Grove Dr., Pasadena, CA 91109, USA}
\altaffiltext{4}{School of Space Research, Kyung Hee University, 1732
Deogyeong-daero, Giheung-gu, Yongin-si, Gyeonggi-do 446-701, Korea}
\altaffiltext{5}{SongAm Space Center, 103, 185 Gwonnyul-ro, Jangheung-myeon, Yangju-si, Gyeonggido 482-812 Korea}
\altaffiltext{6}{Astronomy Program, Department of Physics \& Astronomy, Seoul National University, 1 Gwanak-ro, Gwanak-gu, Seoul 151-742 Korea}
\altaffiltext{7}{Arizona State University, School of Earth and Space Exploration, PO Box 871404, Tempe, AZ 85287-1404, U.S.A.}
\altaffiltext{8}{Korea Astronomy and Space Science Institute, 776 Daedeokdae-ro, Yuseong-gu, Daejeon, Republic of Korea}
\altaffiltext{9}{Subaru Telescope, National Astronomical Observatory of Japan, 650 North A'ohoku Place, Hilo, HI 96720, U.S.A.}
\altaffiltext{$\star$}{Based on observations made with ESO Telescopes at the La Silla Paranal Observatory under programme 091.A-0878.}
\altaffiltext{$\dagger$}{Visiting Astronomer, Kitt Peak National Observatory, National Optical Astronomy Observatory, which is operated by the Association of Universities for Research in Astronomy, Inc. (AURA) under cooperative agreement with the National Science Foundation.}
\altaffiltext{$\ddagger$}{E-mail: ysjeon@astro.snu.ac.kr, mim@astro.snu.ac.kr}

\begin{abstract}

We present our first results of the survey for high redshift quasars at 
$5 \lesssim {\rm z} \lesssim 5.7$.
The search for quasars in this redshift range has been known to be challenging due to limitations of filter sets used in previous studies.  
We conducted a quasar survey for two specific redshift ranges, 4.60 $\leq$ z $\leq$ 5.40 and 5.50 $\leq$ z $\leq$ 6.05, using multi-wavelength data that include observations using
custom-designed filters, $is$ and $iz$.
Using these filters and a new selection technique, 
we were able to reduce the fraction of interlopers. 
Through optical spectroscopy, we 
confirmed
seven quasars at 4.7 $\leq$ z $\leq$ 5.4
with $-27.4 < M_{1450} < -26.4$ which were discovered independently by another group recently. 
We estimated black hole masses and Eddington ratios of four of these quasars 
from optical and near-infrared spectra, and found that these quasars are 
undergoing nearly Eddington-limited accretion which is consistent with the rapid growth of supermassive black holes in luminous quasars at z $\sim$ 5.

\end{abstract}


\section{Introduction}\label{intro}

Observations have shown that large numbers of quasars are found at z $\sim$ 4.5 and at z $> 6$
\citep[e.g.,][]{fan03, fan06131, will07, will10139, jian08, jian09, jian15, mort09, mort11,mcgr13, bana16}.
They harbor supermassive black holes (SMBHs) as massive as $\sim 10^{10}$ M$_{\odot}$ 
\citep[e.g.,][]{jian07, kurk07, jun15, wu15} and 
appear to be vigorously evolving 
\citep{shen11, jian10, im09, jun15}.
However, there 
is a dearth of quasars with measured black hole masses
that makes it difficult 
to investigate how they evolved at 5 $<$ z $<$ 6 
\citep[e.g., Figure 16 in][]{jun15}. Measuring the black hole masses for a significant number of objects at this redshift range allows us to: (1) derive the Eddington luminosities, and consequently, the Eddington ratios, to 
understand the growth of these quasars. 
One simply expect the growth to slow down toward lower redshifts in comparison to z $\sim$ 6;
(2) construct 
the black hole mass function 
to understand the cosmic emergence of the most massive quasars; 
(3) investigate the spectral energy distributions (SEDs) of quasars to 
explore whether 
quasars with very massive black holes have a lower accretion disk temperature
\citep{laor11,wang14}.

\begin{figure*}
\centering
\includegraphics[scale=0.85]{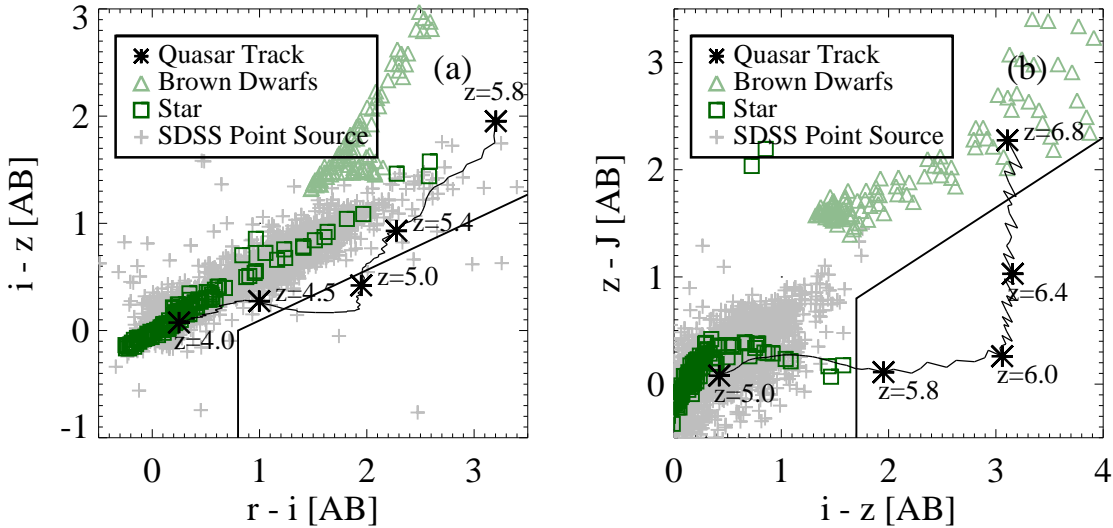}
\caption{
Color-color diagrams adopted by \citet{fan99} (left) and \citet{will09} (right) for high redshift quasar selection. The black solid lines with asterisks are quasar redshift tracks, the triangles are model colors of brown dwarfs, the squares are model colors of stars, and the crosses are point-like sources from the SDSS Star Catalog. The quasar tracks from z = 5.1 to z = 5.7 coincide with late type stars or brown dwarfs. 
The solid boxes indicate the quasar selection boxes.
\label{f_intro_02}}
\end{figure*}

The redshift gap at 5 $<$ z $<$ 6 mentioned above is partly 
due to the inefficiency of quasar selection techniques at 5.2 $<$ z $<$ 5.7 in previous studies
\citep[e.g.,][]{zhen00, shar01, schn01, fan03, fan06131, maha05, cool06, will07, will10139, jian08, jian09, jian15, wu10, iked12, matu13, mcgr13}.
This low efficiency is due to limitations of current filter systems employed by these studies:
the colors of z $\sim$ 5.5 quasars using conventional filters are similar to 
those of late type stars or brown dwarfs. 
Figure \ref{f_intro_02} shows two color-color diagrams generally used for high redshift quasar selection. The black solid lines with asterisks are quasar tracks redshifted from the SDSS composite quasar template from \citet{vand01} including the intergalactic medium (IGM) attenuation \citep{mada96}, the triangles are model colors of brown dwarfs from \citet{burr06}, the squares are model colors of stars by \citet{hewe06} from the Bruzual-Persson-Gunn-Stryker (BPGS) atlas, and the crosses are point-like sources from the Sloan Digital Sky Survey (SDSS) Star Catalog. 
\citet{fan99} used the $r-i$ vs. $i-z$ color-color diagram to identify quasars at z $>$ 4.5 (Figure \ref{f_intro_02}a; $r$-dropout quasars) and \citet{will09} used the $i-z$ vs. $z-J$ color-color diagram for quasars at z $\sim$ 6 (Figure \ref{f_intro_02}b; $i$-dropout quasars). The solid boxes indicate their quasar selection criteria. 
We see that $r$-dropout quasars at z $>$ 5.1 (Figure \ref{f_intro_02}a) and $i$-dropout quasars at z $<$ 5.7 (Figure \ref{f_intro_02}b) are mixed with the late type stars or brown dwarfs on these color-color diagrams. Therefore, the $r$-dropout technique alone cannot be used for z $\sim$ 5.5 quasar selection. As can be seen from above, any configuration of colors from SDSS $ugriz$ or the Two Micron All Sky Survey (2MASS) $JHK$ filters cannot separate quasars at 5.1 $<$ z $<$ 5.7 from stars effectively; a new filter system that exploits the wavelength range between conventional filters is necessary to find these quasars.

\begin{figure}
\centering
\includegraphics[scale=0.4]{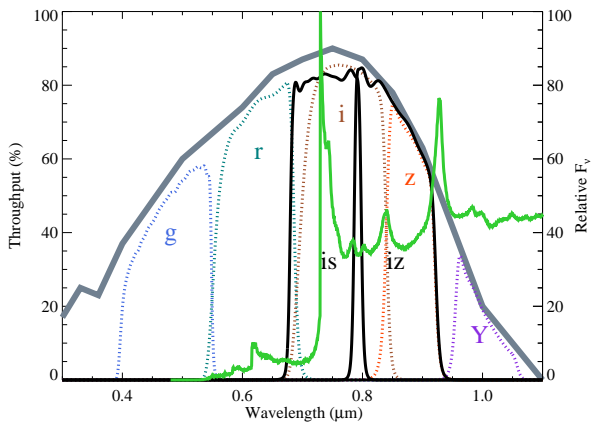}
\caption{
Filter transmission curves of $is$ and $iz$ (black solid lines), SDSS $gri$, and LSST $zY$ bands (colored dashed lines), and the QE of the CCD (gray solid line) of CQUEAN. 
The green line represents the SDSS composite quasar spectrum from \citet{vand01} redshifted to z = 5, 
with IGM attenuation \citep{mada96}.
\label{f_intro_03}}
\end{figure}

Thus, we searched for and studied high redshift quasars at $5 <$ z $< 6$ by using new, additional datasets and performing follow-up observations. 
First, we designed a new filter set, $is$ and $iz$, to supplement the previous filter systems for selecting quasars at this redshift range. Since the central wavelengths of these filters are located between $r$ and $i$, and between $i$ and $z$, respectively, we can select high redshift quasars at this redshift gap, where the SDSS or other filter sets cannot explore.
Second, we needed a special optical detector which has better sensitivity than previous CCDs
at longer wavelengths, leading to more efficient observations with the $is$ and $iz$ filters. Considering these requirements, we developed a CCD camera system, the Camera for QUasars in EArly uNiverse \citep[CQUEAN;][]{kim11, park12, lim13}. 
Equipping a deep-depletion CCD chip to provide high quantum efficiency (QE) at 0.7 -- 1 $\mu$m, we conducted follow-up imaging observations of quasar candidates with the $is$ and $iz$ filters and narrowed down the quasar candidates. CQUEAN was installed on the 2.1-m Otto Struve Telescope at McDonald Observatory in 2010 August, and it has since been used to obtain photometric data for many scientific programs, including our high redshift quasar survey. 
In Figure \ref{f_intro_03}, we plot the filter transmission curves of $is$ and $iz$ (black solid lines), and the SDSS $gri$ and the Large Synoptic Survey Telescope $zY$ bands (colored dashed lines) installed on CQUEAN, with the QE of the CCD taken into consideration (gray solid line). The green line represents the SDSS composite quasar template
redshifted to z $\sim$ 5 and IGM attenuation taken into consideration.  
Note that a similar survey of z $\sim$ 5 luminous quasars is being conducted by \citet{wang16} and \citet{yang16}. Their method relies on the archived multi-wavelength dataset only, while our method includes the use of the custom $is$ and $iz$ filters.

\begin{figure*}
\centering
\includegraphics[scale=0.75, trim={1cm 7.5cm 0 0}]{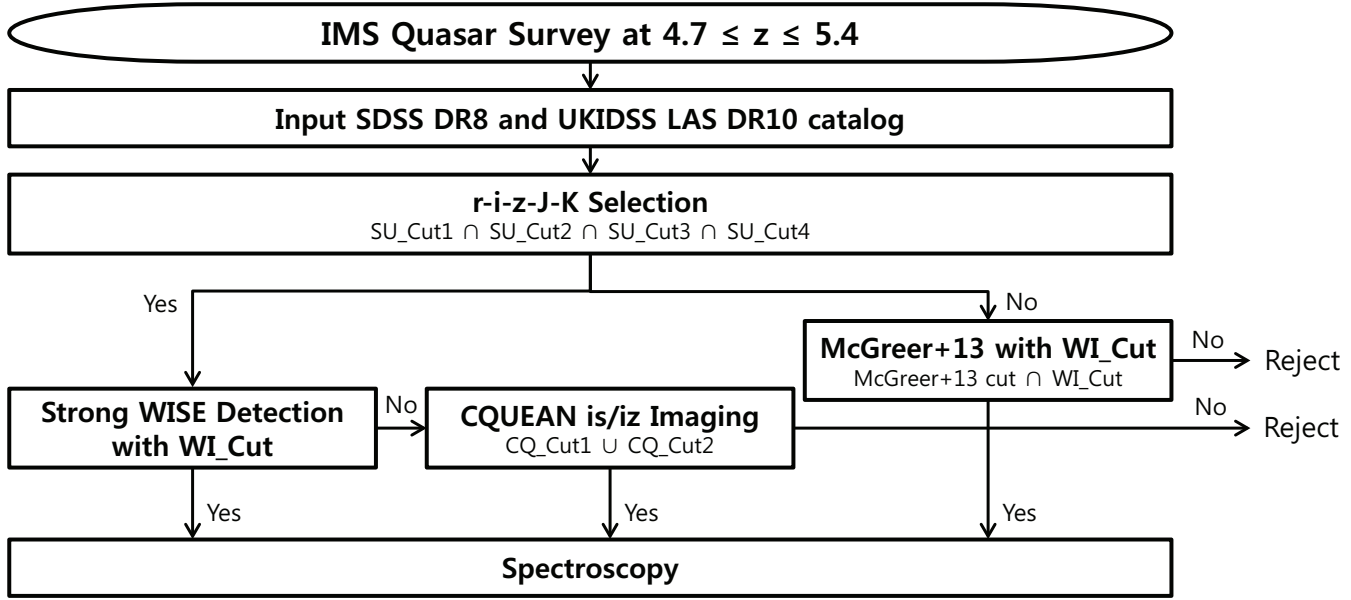}
\caption{
Schematic flow diagram of the main quasar candidate selection algorithm.
\label{f_flowchart}}
\end{figure*}

Section \ref{selec} describes our quasar selection algorithm including color cuts, multi-wavelength data used, and imaging and spectroscopic follow-up observations. 
The photometric and spectroscopic analysis of our discovered quasars are shown in Section \ref{prop}.
We discuss our quasar selection efficiency and expected number of quasars in Section \ref{effi}.
Section \ref{bhmass} presents physical properties of the newly discovered quasars from the spectroscopy.  
We summarize this survey in the final section (Section \ref{summary}). Throughout this paper, we use a cosmology with $\Omega_M=0.3$, $\Omega_\Lambda=0.7$ \citep[e.g.,][]{im97}, and $H_0$ = 70 km s$^{-1}$Mpc$^{-1}$. We use the AB magnitude system.

\section{Quasar Selection and Observation}  \label{selec}

\subsection{Quasar Candidate Selection} \label{selec_ccd}

To select quasars at  $5 <$ z $< 6$ , we employed multi-wavelength data that cover a large area:
SDSS DR8, and the United Kingdom Infra-Red Telescope Infrared Deep Sky Survey Large Area Survey  \citep[UKIDSS LAS;][]{lawr07} DR10; 
the full overlapping area between the two surveys is $\sim$3,400 deg$^2$.
The  $r,$ $i$, $z$, $J$, and $K$ magnitudes are used. 
Since the contamination rate using these filters is still high, 
we adopted 
$is$ and $iz$-band photometry to discriminate brown dwarfs from $r$-dropout objects. 
Then we set additional criteria to assign priorities for follow-up observations.
No stellarity cut is made to avoid missing quasars that are classified to be extended objects (e.g., due to
host galaxy or noise in stellarity calculation), although we used the stellarity as a way to set priorities for
follow-up observation.
Figure \ref{f_flowchart} shows a main quasar candidate selection algorithm.

\subsubsection{$r-i-z-J-K$, \textit{is} and \textit{iz}-band Selections}\label{selec_sduk}

\begin{figure}
\centering
\includegraphics[scale=0.55]{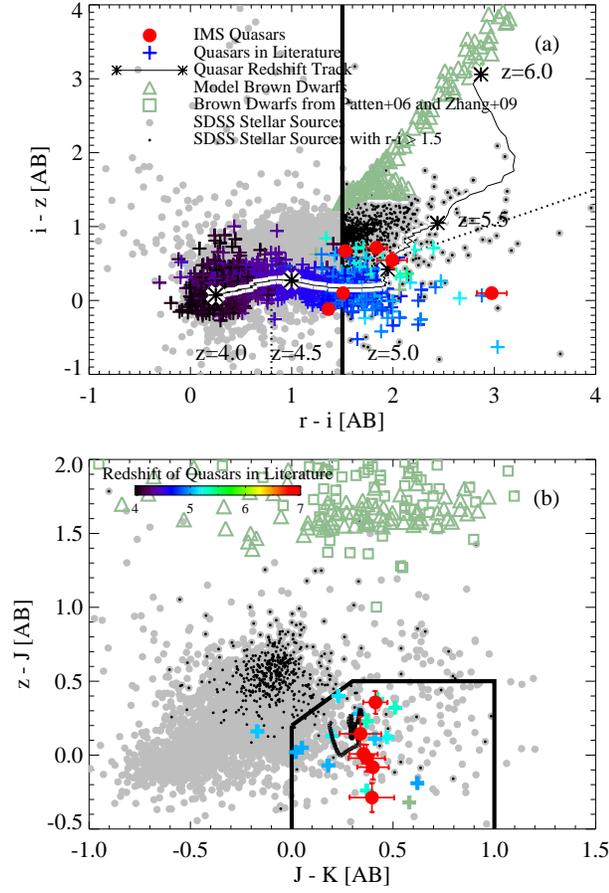}
\caption{
Two color-color diagrams we adopted for quasar selection at $5 \lesssim {\rm z} \lesssim 5.7$.
We plot the model brown dwarfs (green triangles), observed brown dwarfs from \citet{patt06} and \citet{zhan09} (green squares), stellar sources from SDSS (gray circles), previously discovered quasars (crosses), 
and the redshift tracks of quasars at 4 $<$ z $<$ 6 (black solid line with asterisks in (a) and black solid line in (b)).
The thick solid lines indicate the boxes for our quasar selection and the dotted box in (a) is the selection box of \citet{fan99} for comparison.
The black dots are SDSS stellar sources with $r-i > 1.5$,
showing a high contamination rate even after the $z-J-K$ cut.
We plotted our 6 new quasars with red circles (this work)
and most of them are within the selection boxes. One exception is IMS J0324+0426 in the $r-i-z$ color-color diagram, which was selected using the color cuts of \citet{mcgr13}.
\label{f_selec_02}}
\end{figure}

To select quasar candidates from broadband photometry, we used the dropout feature at the
Lyman $\alpha$ (Ly$\alpha$) emission line that are common in high redshift objects. 
The Ly$\alpha$ dropouts can be identified using the $r-i$ color for quasars at z $>$ 3.6, and $r-i$ $>$ 1.5 for quasars at z $>$ 4.6. 
To discriminate high redshift quasars from red, low mass stars,
we used 
three color cuts, $r-i$, $z-J$, and $J-K$: $r-i$ to select dropout objects, $z-J$ to remove  brown dwarfs, and $J-K$ to eliminate other stars. 
Figure \ref{f_selec_02} shows two color-color diagrams with model brown dwarfs from \citet{burr06} (green triangles), observed brown dwarfs from \citet{patt06} and \citet{zhan09} (green squares), and stellar sources from the SDSS catalog (gray circles, $\sim$10,000 randomly selected sources). The black dots indicate SDSS stellar sources with $r-i > 1.5$.
To verify the position of quasars at 4 $<$ z $<$ 6, we plotted previously discovered quasars from the SDSS DR7 quasar catalog and \citet{leip14}  (crosses; the color indicates its redshift, as shown on the color bar in Figure \ref{f_selec_02}b). The quasar redshift track at 4 $<$ z $<$ 6 is plotted with the black solid line by assuming 
the redshifted and IGM-attenuated SDSS composite quasar template.
The thick solid lines indicate the selection cuts for our quasar selection and the dotted box in (a) is the selection box from \citet{fan99} for comparison.
The selection boxes from SDSS and UKIDSS LAS datasets are defined as below:
\begin{quote}
\texttt{SU\_Cut1}) $r-i > 1.5$ \\
\texttt{SU\_Cut2}) $[0 < J-K < 1] ~\cap~ [-1 < z-J < 0.5] ~\cap~ [(z-J) < (J-K)+0.2]$
\end{quote}  
\texttt{SU\_Cut1} is for selecting the $r$-dropout objects and \texttt{SU\_Cut2} is for weeding out late type stars and brown dwarfs. 
Since \texttt{SU\_Cut1} does not adopt the $i-z$ cut, unlike \citet{fan99}, quasar candidates at z $\sim$ 5.5 can be selected with this color cut.
However, since the selection box of \texttt{SU\_Cut2} is close to the stellar locus (gray circles) and part of the stellar sources selected from \texttt{SU\_Cut1} is still located inside \texttt{SU\_Cut2} (black circles inside \texttt{SU\_Cut2}),
the selected sample is still significantly contaminated by stars (more than 99\% of the selected objects are expected to be stars; see Section \ref{effi_numden}).

To reduce stellar contamination in our sample,  
we impose magnitude cuts in the shorter wavelength data, as well as in the $z$-band. 
We set the magnitude cuts as below: 
\begin{quote}
\texttt{SU\_Cut3}) $u$, $g$ fainter than the 3$\sigma$ detection limits ($u > 22.85$ and $g > 23.55$ mag)\\
\texttt{SU\_Cut4}) $z < 19.5$ mag
\end{quote}  
\noindent
From the cross-matched sources 
from SDSS DR8 and UKIDSS LAS DR10, 98.4\% of sources are rejected via the above four criteria, and,
after visual inspection for false detection, about 3,600 candidates are finally listed.
We checked the sources which were classified as quasars at z $>$ 4.6 from the SDSS DR7 quasar catalog and found that 
14 quasars at 4.69 $<$ z $<$ 5.29 and 2 quasars at 5.50 $<$ z $<$ 6.05 were already spectroscopically identified.
These $\sim$3,600 candidates still contain a significant fraction of contaminants considering that the expected number of quasars at z $\sim$ 5 in 3,400 deg$^2$ is $\sim$30 (Section \ref{effi_numden}), 
showing that about 99\% of these sources will be interlopers.    
This is because the selected candidates from these two color-color diagrams are still contaminated by stellar sources,
which are shown as the black circles inside \texttt{SU\_Cut2} in Figure \ref{f_selec_02}b.
To eliminate these contaminants, we employed an additional selection method: 
photometry from $is$/$iz$-bands. 

\begin{figure}
\centering
\includegraphics[scale=0.6]{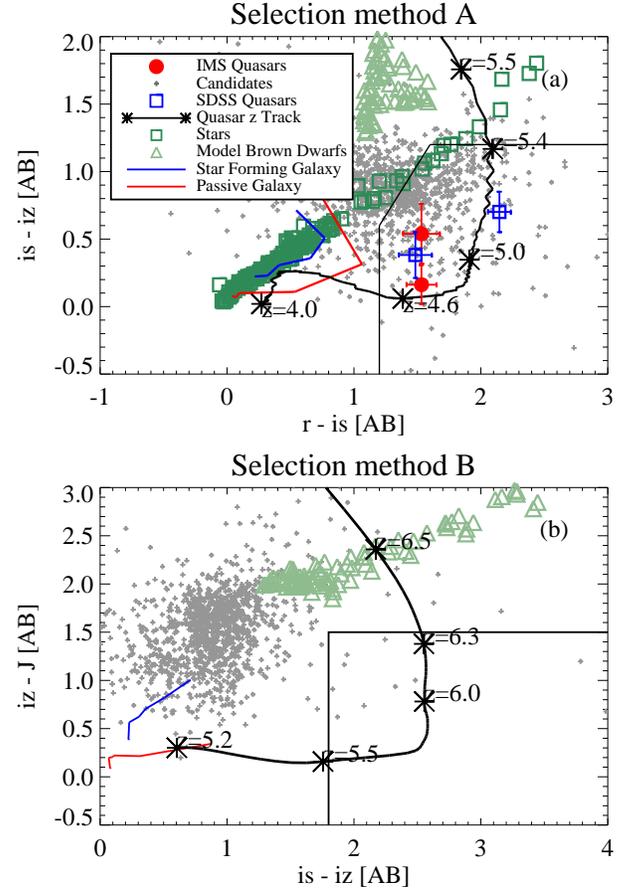}
\caption{
Two color-color diagrams using $is$ and $iz$-bands. 
Quasar candidates (gray crosses), 
SDSS quasars (blue squares), 
quasar redshift tracks (black lines with asterisks),
model brown dwarfs (green triangles),
stars (green squares), 
star forming galaxy redshift tracks (blue lines), 
passive galaxy redshift tracks (red lines), 
and the two selection boxes are plotted.  
We plotted our two new quasars with the $is$ and $iz$ photometry using red circles in (a). 
\label{f_selec_03}}
\end{figure}

\begin{figure}
\centering
\includegraphics[scale=0.45]{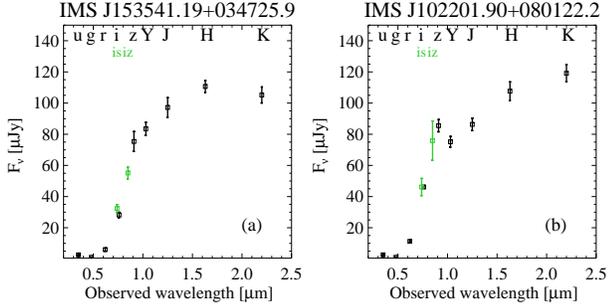}
\caption{
Examples of SEDs of $u, g, r, is, i, iz, z, Y, J, H,$ and $K$-bands. 
The filter names are marked at each wavelength. 
The $is$ and $iz$ filters are plotted with green points.
(a): A candidate with blue $H-K$ color ($H-K=-0.06$). 
(b): A candidate which turned out to be a high redshift quasar ($H-K=0.11$).  
\label{f_selec_07}}
\end{figure}

We now apply selection cuts
using the $is$ and $iz$-bands of CQUEAN. 
The color cuts were defined using quasar redshift tracks.
We optimized our quasar selection 
using \texttt{CQ\_Cut1} ($r-is-iz$: selection method A) or \texttt{CQ\_Cut2} ($is-iz-J$: selection method B) on color-color diagrams,
which explore the redshift ranges of 4.60 $\leq$ z $\leq$ 5.40 and 5.50 $\leq$ z $\leq$ 6.05, respectively (Section \ref{effi_color}).
The criteria for the selections are: 
\begin{quote}
\texttt{CQ\_Cut1} ($r-is-iz$ for 4.60 $\leq$ z $\leq$ 5.40): selection method A\\
$~~~~~~~~~[r-is > 1.2] ~\cap~ [is-iz < 1.2] ~\cap~ [is-iz < 1.5 \times (r-is)-1.2]$\\
\texttt{CQ\_Cut2} ($is-iz-J$ for 5.50 $\leq$ z $\leq$ 6.05): selection method B\\
$~~~~~~~~~[s-iz > 1.8] ~\cap~ [iz-J < 1.5]$.
\end{quote}
Figure \ref{f_selec_03} shows these two color-color diagrams with 
quasar redshift tracks 
(black lines with asterisks; from the redshifted and IGM-attenuated SDSS composite quasar template), 
model brown dwarfs 
\citep[green triangles; from][]{burr06},
stars from \citet{gunn83} (green squares), 
star forming galaxy redshift tracks (blue line; model colors from M51), passive galaxy redshift tracks (red line; model colors from the \citet{bruz03} model 
of a passively evolving 5 Gyr-old galaxy with spontaneous burst, metallicity of Z = 0.02, and  the Salpeter initial mass function), 
and SDSS quasars with $is$ and $iz$ observations for comparison (blue square). 
The two color cuts are denoted. 
About 1,400 among $\sim$3,600 quasar candidates were imaged with CQUEAN (gray crosses) 
and among them,  about 500 candidates satisfy these color cuts.
However, selected candidates in \texttt{CQ\_Cut1} still show a high contamination rate 
because the stellar locus is found near the quasar redshift track. 
After considering the spectral shape of quasars, we selected about 60 targets as promising candidates 
via visual inspection of SEDs, because quasars at 5 $<$ z $<$ 6 tend to have $H-K$ colors redder 
than those of dwarf stars ($H-K \gtrsim 0$) due to the power-law continuum of quasars.
During the visual inspection, SEDs that show a turn down in flux toward longer wavelengths (Figure \ref{f_selec_07}a) are rejected in comparison to those that are retained as candidates (Figure \ref{f_selec_07}b).

\subsubsection{Ancillary Selection} \label{selec_cri}

\begin{figure}
\centering
\includegraphics[scale=0.55]{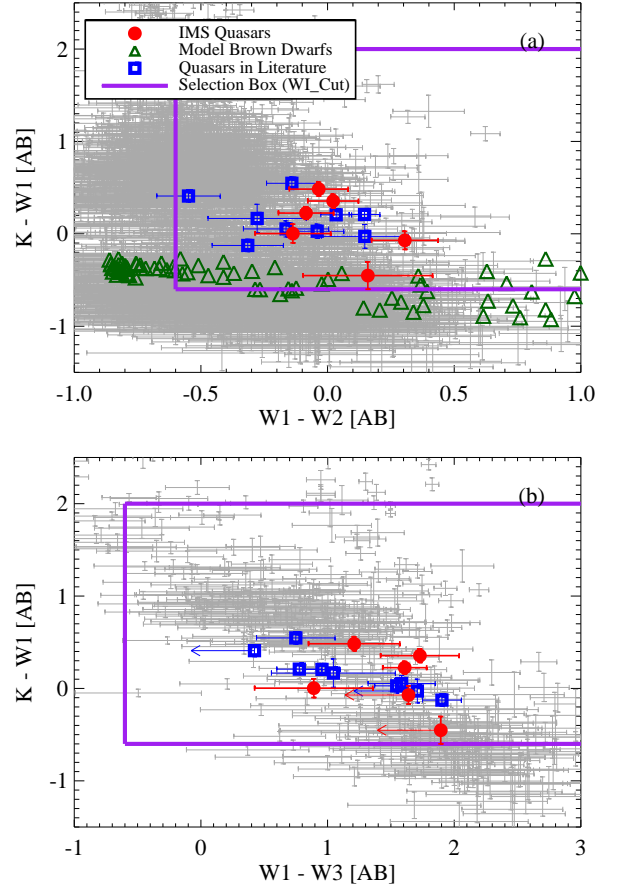}
\caption{
Two color-color diagrams with $WISE$ photometry and our selection boxes. 
We plot our $\sim$3,600 candidates with $WISE$ detections (gray crosses), previously known z $\sim$ 5 quasars (blue squares), and model brown dwarfs (green triangles).
We plotted our 6 new quasars with red circles.
\label{f_selec_06}}
\end{figure}

We set additional selection criteria for assigning priorities 
for imaging and spectroscopic follow-up observations.

\textbf{WISE Selection:}
The $WISE$ catalog 
provides 3.4, 4.6, and 12 micron data ($W1$, $W2$, and $W3$-bands) that are useful for quasar candidate selection: due to the nature of quasar continua, 
we expect quasars at z $\sim$ 5 to have $-0.6 < K-W1 < 2.0$ and $W1-W2 > -0.6$ while about 60\% of brown dwarfs do not.
The cut of $W1-W3 > -0.6$ is also adopted to remove the brown dwarf outliers, although this cut is not
as powerful as the other WISE cuts.
We selected red sources in $WISE$ bands and assigned high priorities to these sources for follow-up observations. 
Figure \ref{f_selec_06} shows our $\sim$3,600 candidates with $WISE$ detections (gray crosses), 9 previously discovered quasars 
with $WISE$ detections (blue squares), and model brown dwarfs (green triangles).
Since the model brown dwarf templates from \citet{burr06} do not extend to the $W3$-band, only Figure \ref{f_selec_06}a shows the colors of model brown dwarfs (green triangles).
We do not consider the quasar redshift track since the rest-frame optical region of the quasar template from \citet{vand01}, which are sampled by WISE bands, are affected by host galaxy. 
Therefore, based on the observed quasar colors, 
we defined \texttt{WI\_Cut} (purple boxes). 
We adopt the following selections: 
\begin{quote}
\texttt{WI\_Cut}: $[W1-W2 > -0.6]$ $\cap$ $[-0.6 < K-W1 < 2]$\\ 
$~~~~~~~~$and/or $[W1-W3 > -0.6]$ $\cap$ $[-0.6 < K-W1 < 2]$ 
\end{quote}
\noindent
Candidates detected in WISE bands were assigned higher priorities and
some of them showing strong power law continuum at the rest-frame ultraviolet spectral region 
were followed-up with optical spectroscopy. 
53 candidates were given higher priorities due to the WISE criteria (see Table \ref{t_pri}).

\begin{deluxetable*}{c|c|c|c|r}
\tabletypesize{\normalsize}
\tablecaption{Priorities for CQUEAN imaging follow-up observations \label{t_pri}}
\tablewidth{0pt}
\tablehead{
Priority&Stellarity&WISE&McGreer+13 or Polsterer+13& Number
}
\startdata

0&yes&yes&yes&8\\
\hline
1&yes&no&yes&24\\
\hline
2&yes&yes&no&45\\
\hline
3&yes&no&no&1,039\\
\hline
4&\multicolumn{3}{c|}{yes or yes or yes}&1,142\\
\hline
5&no&&yes&123\\
\hline
10&\multicolumn{3}{c|}{others}&1,105

\enddata

\end{deluxetable*}

\textbf{Color cuts from \citet{mcgr13}:}
\citet{mcgr13} discovered a number of quasars at 4.7 $<$ z $<$ 5.1  over the area covered by SDSS, including Stripe 82.  
From the cross-matched sources from SDSS DR8 and UKIDSS LAS DR10, 
148 candidates with $z < 19.5$ mag  satisfy the these conditions 
and 9 of them are included in our $\sim$3,600 quasar candidates.
We gave high priorities to our candidates that satisfied the color cuts used in their work. 
Sources selected from these color cuts with $WISE$ selection, but not included in the $r-i-z-J-K$ color cuts, are also added to our candidate list.

\textbf{Candidates from \citet{pols13}:}
\citet{pols13} provide a quasar candidate catalog containing 121,909 sources with their photometric redshifts at 2.558 $\leq$ z $\leq$ 6.131. 10 sources are included in our candidate list and we gave higher priorities to these sources.

\textbf{Stellarity:}
We use \texttt{mergedClass} for UKIDSS LAS and \texttt{type} for SDSS to distinguish point sources from extended sources. We defined that a source with \texttt{mergedClass} = $-1$ or $-2$, or \texttt{type} = 6, is a point source, and gave higher priorities to these sources. We did not exclude the extended sources because 17\% of the discovered quasars from \citet{mcgr13} are classified as extended sources in their $i$-band, meaning that some quasars may be classified as extended sources.

\subsubsection{Selection Summary} \label{selec_sum}

The selection method used in this paper can be summarized as the following. We begin with an adjoint sample of SDSS DR8 and UKIDSS LAS DR10. We select objects showing Ly$\alpha$ drops between $r$ and $i$, and remove brown dwarfs and stars using the $r-i-z-J-K$ color-color diagrams (\texttt{SU\_Cut1,2}). To decrease the number of stellar contaminants, we adopt magnitude cuts in the $u$, $g$, and $z$ bands (\texttt{SU\_Cut3,4}). These four criteria decrease the sample to $\sim$3,600 objects. 
Among them, sources with strong WISE detection and WISE selection (\texttt{WI\_Cut}) are listed as promising candidates. 
Objects not included  
in the $r-i-z-J-K$ selection, but selected from the \citet{mcgr13} cuts with WISE selection (\texttt{WI\_Cut}), are added to the candidate list.
Among the $\sim$3,600 candidates, to reduce contamination, we utilized two color-color diagrams, $r-is-iz$ and $is-iz-J$, employing our new filter system and selected quasar candidates at two redshift ranges (\texttt{CQ\_Cut1,2}).
For the CQUEAN imaging follow-up observations, we set priorities of our candidates considering the stellarity, the WISE detection, the color cuts from \citet{mcgr13}, and candidates from \citet{pols13}. 
Objects showing point-like shapes with WISE detections as well as satisfying the color cuts from \citet{mcgr13}
or candidate list from \citet{pols13} were classified as the important candidates. 
Table \ref{t_pri} lists the priority for each case, with smaller numbers indicating higher priorities.
We have been conducting the $is$ and $iz$ imaging for the high priority objects and about half of the sample was imaged in these two filters.  
Finally, via visual inspection of the SEDs, $\sim$60 targets were selected to be our main samples for spectroscopy.

\subsection{Optical Imaging Follow-up Observations with CQUEAN}  \label{obs_opimg}

Follow-up observations of our high redshift quasar candidates using CQUEAN began in 2010 August and are still on-going. About 1,400 among $\sim$3,600 candidates with high priorities have been observed with CQUEAN until now. 

We used short single exposure times of 30 sec for $iz$ and 60 sec for $is$ filters, respectively. 
Number of frames varied 
depending on the sky conditions, such as seeing conditions and extinction. If the peak value of a target was greater than 80 ADU after a 30 sec exposure with $iz$, 2.5 (30 sec $\times$ 5) and 5 (60 sec $\times$ 5) minutes were used as the integration times for the  $iz$ and $is$ filters, respectively. If the signal was lower than the criterion, we exposed 5 (30 sec $\times$ 10) and 10 (60 sec $\times$ 10) minutes with $iz$ and $is$, respectively, or more.  

Preprocessing including bias subtraction, dark subtraction and flat fielding, were preformed using the usual data reduction procedures in the IRAF\footnote{IRAF is distributed by the National Optical Astronomy Observatory, which is operated by the Association of Universities for Research in Astronomy,  Inc.,  under cooperative agreement with the National Science Foundation.} \texttt{noao.imred.ccdred} package. Since the bias values may change with time \citep{park12}, we used bias images that 
were taken closest to the object frames, time-wise. 
We combined images of each field and  filter in average. We used the \texttt{ccmap} task of IRAF and SCAMP \citep{bert06} to derive astrometric solutions. SExtractor 
\citep{bert96} was used for the source detection and photometry. We derived auto-magnitudes which are taken as the total magnitudes.
 
For the photometric calibration, we used SDSS photometry of stellar objects
inside each target field.
We performed $\chi^2$ fitting to the SDSS $r$, $i$, $z$ magnitudes of stellar sources, to determine best-fit stellar spectral types.
For this, we used the SED 
templates from \citet{gunn83}, containing 175 spectra of various stellar types. 
The model $is$ and $iz$ magnitudes were calculated from the best-fit templates and these are used to define the 
zero-points ($Zp$) of each filter image of each field. 
The $Zp$ values were calculated for each star, and we took the average of these values as 
$Zp$ and the standard deviation of the scatters as its $Zp$ error. The average $Zp$ error is about 0.05 mag. During the calculation, objects with large reduced $\chi^2$ values ($\chi^2_{\nu}$ $>$ 5) were rejected for the estimation. 
Note that this photometric calibration method is described in more detail in \citet{jeon16}.

\begin{deluxetable*}{clcccccccccccc}
\tabletypesize{\scriptsize}
\tablecaption{Spectroscopic observation summary of IMS quasars\label{t_spec}}
\tablewidth{0pt}
\tablehead{
Spectroscopy&Date&Telescope&Target&Integration Time (min)&Slit Width ($\arcsec$)}

\startdata

Optical& 2013 Jan. 16&KPNO 4-m&IMS J1022+0801&80&3.0\\
& 2013 May 6&NTT&IMS J1437+0708&40&1.2\\
& 2013 May 6&NTT&IMS J2225+0330&90&1.0\\
& 2013 May 7&NTT&IMS J1437+0708&60&1.0\\
& 2013 Sep. 27&KPNO 4-m&IMS J0122+1216&45&1.5\\
& 2013 Sep. 28&KPNO 4-m&IMS J0155+0415&60&1.5\\
& 2013 Sep. 28&KPNO 4-m&IMS J0324+0426&45&1.5\\
& 2013 Sep. 29&KPNO 4-m&IMS J2225+0330&60&1.5\\
& 2013 Sep. 29&KPNO 4-m&IMS J0122+1216&45&1.5\\
\hline
NIR&  2014 Oct. 6&Magellan&IMS J0122+1216&60&1.0\\ 
& 2014 Oct. 7&Magellan&IMS J0155+0415&30&1.0\\
& 2014 Oct. 6&Magellan&IMS J0324+0426&60&1.0\\
& 2015 Aug. 30&Gemini-N&IMS J2225+0330&53&0.675
\enddata
\end{deluxetable*}

\subsection{Optical Spectroscopic Follow-up Observations} \label{obs_opspec}

We observed 47 candidates using the Kitt Peak National
Observatory (KPNO) 4-m Mayall telescope and the European Southern Observatory (ESO) New Technology Telescope (NTT). The KPNO 4-m observations were performed over three runs
for 10 nights from 2013 January to September, and the NTT observation was done for 3 nights in 2013 May.
 
For the observations at KPNO, we used the  
Ritchey-Chr\'{e}tien Focus Spectrograph in a longslit mode (RCSPL\footnote{http://www-kpno.kpno.noao.edu/manuals/l2mspect/index.html}) with a LB1A CCD, the BL400 grating of R $\sim$ 500 for a 2$\arcsec$ slit, and OG400 filter. 
LB1A uses a thick CCD chip, therefore it does not suffer much from fringing.
The wavelength coverage is 5,000$\AA$ -- 10,000$\AA$.
For the observation at the ESO NTT, we used the ESO Faint Object Spectrograph and Camera v.2 \citep[EFOSC2; ][]{buzz84}.
The EFOSC2 was used with Gr\#2 that has a wavelength coverage of 5,100$\AA$ -- 11,000$\AA$ and R $\sim$ 135 for a 1$\arcsec$ slit.
We took calibration frames including bias, dark, flat, and arc. Standard stars such as G191B2B, GD153, CD-32d9927, LTT7379, LTT3864, Feige110, and HR7596 were observed for the flux calibration. The slit widths varied from $1\farcs0$ to $3\farcs0$, depending on the seeing conditions. 
Table \ref{t_spec} shows the 
summary of the optical spectroscopic observations of the discovered quasars, namely the total integration time and the slit width for each target.

We followed the typical steps for preprocessing, 
including bias subtraction, dark subtraction, and flat fielding, for each science image, standard star image and arc image, using the \texttt{noao.imred.ccdred} package in IRAF. The spectra were extracted using the \texttt{noao.imred.kpnoslit} or the \texttt{noao.twodspec.apextract} packages in IRAF for each single image. We used an optimal aperture size for each image where the S/N is highest. After this, wavelength and flux calibrations were conducted. The spectra were flux-calibrated using spectra of the standard stars.
Considering the light loss due to variable seeing conditions, we scaled the spectra using broadband photometry. 
We chose $i$-band for this calibration, since we get  
the highest S/N in this band  for the observed spectra. 
The flux-calibrated spectra were combined in median using the \texttt{scombine} task of IRAF and  
were corrected for Galactic extinction using values from
\citet{card89} and \citet{schl98}.

We observed 47 candidates and 6 of them turned out to be high redshift quasars at 4.7 $\leq$ z $\leq$ 5.4: these are referred to as Infrared Medium-deep Survey (IMS) quasars.
Table \ref{t_imsq} lists the names, coordinates, and redshifts (Section \ref{prop_phys}) of the 6 quasars. 
The naming convention of our quasars is IMS JHHMMSS.SS$\pm$DDMMSS.S in J2000.0 coordinates 
(IMS JHHMM$\pm$DDMM for brevity).

\begin{deluxetable*}{cclccccc}
\tabletypesize{\scriptsize}
\tablecaption{General information of IMS quasars \label{t_imsq}}
\tablewidth{0pt}
\tablehead{
Name&R.A. and Dec. (J2000.0) &Redshift&$M_{1450}$
}
\startdata
IMS J032407.70+042613.3&03:24:07.70+04:26:13.3&4.70(Ly$\alpha$)\tablenotemark{a}, 4.68(\civ\,), 4.73(\mgii\,)& $-$27.21$\pm$0.29\\
IMS J012247.33+121623.9&01:22:47.33+12:16:23.9&4.83(Ly$\alpha$)\tablenotemark{b}, 4.81(\civ\,)& $-$26.47$\pm$0.68\\
IMS J143704.82+070808.3&14:37:04.82+07:08:08.3&4.94(Ly$\alpha$)\tablenotemark{c} & $-$27.14$\pm$0.09\\
IMS J222514.39+033012.6&22:25:14.39+03:30:12.6&5.35(Ly$\alpha$)\tablenotemark{d},\hspace{1.5cm} 5.26(\mgii\,)& $-$26.47$\pm$0.29\\
IMS J102201.90+080122.2&10:22:01.90+08:01:22.2&5.36(Ly$\alpha$)& $-$27.38$\pm$0.10\\
IMS J015533.28+041506.8&01:55:33.28+04:15:06.8&5.35(Ly$\alpha$)\tablenotemark{e}, 5.27(\civ\,)& $-$26.85$\pm$1.09

\enddata
\tablenotetext{a}{z${\rm _{spec}}$=4.72 from \citet{wang16}}
\tablenotetext{b}{z${\rm _{spec}}$=4.76 from \citet{yi15} and z${\rm _{spec}}$=4.79 from \citet{wang16}}
\tablenotetext{c}{z${\rm _{spec}}$=4.93 from \citet{wang16}}
\tablenotetext{d}{z${\rm _{spec}}$=5.24 from \citet{wang16}}
\tablenotetext{e}{z${\rm _{spec}}$=5.37 from \citet{wang16}}
\tablecomments{z${\rm _{spec}}$ from other papers are all derived from Ly$\alpha$}
\end{deluxetable*}

\subsection{NIR Spectroscopic Observation} \label{obs_nirspec}

To measure their black hole masses and Eddington ratios,  
we observed four of the six newly discovered quasars 
using the Folded-port InfraRed Echellette (FIRE\footnote{http://web.mit.edu/$\sim$rsimcoe/www/FIRE/index.html}) spectrograph on the Magellan telescope
(IMS J0324+0426, IMS J0122+1216, and IMS J0155+0415)
and using the Gemini Near Infra-Red Spectrograph (GNIRS) 
on the Gemini North (Gemini-N) telescope
(IMS J2225+0330; program GN-2015B-Q-77). 
Table \ref{t_spec} shows the summary of the Magellan and Gemini-N observations. 

In the Magellan/FIRE observation, 
we used a slit width of $1\farcs00$ with the Echelle mode (R = 3,600). The ABBA pointing method was used for the sky subtraction between exposures. We observed standard stars for each target. Data for the flat fielding and the wavelength calibration were also taken. 
The data reduction was conducted using the IDL suite, \texttt{FIREHOSE}. This pipeline conducts the preprocessing, object extraction, telluric correction, flux calibration, and spectra combining. 

In the Gemini-N/GNIRS observation, 
we used the cross-dispersed (XD) mode with the 32 line mm$^{-1}$ grating, 
the short blue camera, and its SXD prism.
Adopting the slit of $0\farcs675$ width, 
we obtained R $\sim$ 800.
We also used the ABBA pointing method and observed standard stars and calibration data. 
For the data reduction, we use the Gemini IRAF package
following the reduction scripts in the Gemini web page\footnote{https://www.gemini.edu/sciops/instruments/gnirs/data-format-and-reduction/reducing-xd-spectra}. 
The steps include pattern noise cleaning using the \texttt{clearnir} script,
reducing the science data using flatfield images, combining images, wavelength calibration, 
extracting spectra, and flux calibration using standard stars.

We scaled the flux of the combined spectra using broadband photometry. 
After that, the spectra were corrected for Galactic extinction using \citet{card89} and \citet{schl98}.

\section{High Redshift Quasars} \label{prop}

\subsection{Photometric Properties }\label{prop_phot}

\begin{deluxetable*}{cccccccccccccccc}
\tabletypesize{\scriptsize}
\tablecaption{Optical photometric information of IMS quasars\label{t_optp}}
\tablewidth{0pt}
\tablehead{
Name	&	$g$	&	$r$	&	$i$	&	$z$	&	$is$&$iz$
}
\startdata
IMS J0324+0426	&	23.95$\pm$0.39	&	20.39$\pm$0.04	&	19.03$\pm$0.03	&	19.15$\pm$0.06	&\nodata & \nodata\\
IMS J0122+1216	&	24.29$\pm$0.37	&	22.35$\pm$0.14	&	19.37$\pm$0.03	&	19.27$\pm$0.06	&\nodata & \nodata\\
IMS J1437+0708	&	25.02$\pm$0.72	&	20.71$\pm$0.04	&	19.20$\pm$0.02	&	19.10$\pm$0.06	&19.17$\pm$0.11&19.01$\pm$0.09\\
IMS J2225+0330	&	25.67$\pm$0.68	&	22.01$\pm$0.14	&	20.02$\pm$0.05	&	19.47$\pm$0.10	&\nodata & \nodata\\
IMS J1022+0801	&	25.23$\pm$0.64	&	21.27$\pm$0.06	&	19.74$\pm$0.02	&	19.07$\pm$0.05	&19.74$\pm$0.13&19.20$\pm$0.18\\
IMS J0155+0415	&	24.07$\pm$0.38	&	21.81$\pm$0.10	&	19.98$\pm$0.03	&	19.26$\pm$0.06	&\nodata & \nodata

\enddata
\end{deluxetable*}

\begin{deluxetable*}{cccccccccccccccc}
\tabletypesize{\tiny}
\tablecaption{NIR photometric information of IMS quasars\label{t_nirp}}
\tablewidth{0pt}
\tablehead{
Name&$W1$&$W2$&$W3$&$W4$&$Y$	&	$J$	&	$H$	&	$K$ 	
}
\startdata
IMS J0324+0426&18.47$\pm$0.05&18.45$\pm$0.09&16.74$\pm$0.31&15.27$\pm$0.38&19.39$\pm$0.05&19.23$\pm$0.05&18.96$\pm$0.05&18.83$\pm$0.05\\
IMS J0122+1216&18.28$\pm$0.05&18.36$\pm$0.09&16.67$\pm$0.17&99.00$\pm$99.00&19.12$\pm$0.04&18.92$\pm$0.04&18.56$\pm$0.04&18.50$\pm$0.04\\
IMS J1437+0708&18.99$\pm$0.07&19.12$\pm$0.13&18.10$\pm$0.46&99.00$\pm$99.00&19.40$\pm$0.05&19.39$\pm$0.08&19.01$\pm$0.06&18.99$\pm$0.08\\
IMS J2225+0330&19.44$\pm$0.12&19.28$\pm$0.22&99.00$\pm$99.00&99.00$\pm$99.00&19.48$\pm$0.06&19.33$\pm$0.06&19.04$\pm$0.10&18.99$\pm$0.08\\
IMS J1022+0801&18.23$\pm$0.05&18.26$\pm$0.10&17.01$\pm$0.36&99.00$\pm$99.00&19.21$\pm$0.05&19.06$\pm$0.05&18.82$\pm$0.06&18.71$\pm$0.05\\
IMS J0155+0415&18.98$\pm$0.08&18.68$\pm$0.11&99.00$\pm$99.00&99.00$\pm$99.00&19.66$\pm$0.07&19.28$\pm$0.06&19.00$\pm$0.06&18.91$\pm$0.06
\enddata
\tablecomments{We used a dummy value of 99.99 for non-detections.}
\end{deluxetable*}

\begin{deluxetable*}{cccccccccccccccccc}
\tabletypesize{\scriptsize}
\tablecaption{Selection methods of IMS quasars \label{t_prop}}
\tablewidth{0pt}
\tablehead{
Name&	WISE\tablenotemark{a} 	&	WISE\tablenotemark{b}	&	McGreer+13\tablenotemark{c}	&	Polsterer+13\tablenotemark{d} &$r-is-iz$\tablenotemark{e}&$is-iz-J$\tablenotemark{f}	\\
	    &	($K-W1-W2$)	&	($K-W1-W3$)	&				& &&
}
\startdata

IMS J0324+0426	&	yes	&	yes			&	yes	&	no	&\nodata&\nodata\\
IMS J0122+1216	&	yes	&	yes			&	yes	&	no	&\nodata&\nodata\\
IMS J1437+0708	&	yes	&	yes			&	yes	&	yes 	&yes&\nodata\\
IMS J2225+0330	&	yes	&	yes			&	no	&	no	&\nodata&\nodata\\
IMS J1022+0801	&	yes	&	yes			&	no	&	no	&yes&\nodata\\
IMS J0155+0415	&	yes	&	yes			&	no	&	no	&\nodata&\nodata

\enddata

\tablenotetext{a}{
Does it satisfy the color cut of $K-W1-W2$?
}
\tablenotetext{b}{
Does it satisfy the color cut of $K-W1-W3$?
}
\tablenotetext{c}{
Does it satisfy the color cuts from \citet{mcgr13}?
}
\tablenotetext{d}{
Is it contained in the candidate list from \citet{pols13}?
}
\tablenotetext{e}{
Does it satisfy the color cut of $r-is-iz$?
}
\tablenotetext{f}{
Does it satisfy the color cut of $is-iz-J$?
}

\end{deluxetable*}

We list the photometric information from SDSS, UKIDSS LAS, $WISE$, and CQUEAN of our newly discovered quasars in  Tables \ref{t_optp} and \ref{t_nirp}. 
Table \ref{t_prop} shows their selection properties. 
All six of them have $WISE$ detections and are located inside the $WISE$ color cuts (Figure \ref{f_selec_06}; $K-W1-W2$ or $K-W1-W3$).
IMS J0324+0426, IMS J0122+1216, and IMS J1437+0708
also satisfy the color cuts of \citet{mcgr13} that are aimed at selecting 
z $<$ 5.1 quasars. \citet{pols13} provided a photometric redshift for IMS J1437+0708 of z=4.961$\pm$0.127, which is in agreement with our redshift measured from the Ly$\alpha$ emission line (Section \ref{prop_phys}). 
For the two IMS quasars with $is$ and $iz$ photometry,  
Figure \ref{f_selec_03}a shows their colors in the $r-is-iz$ color-color diagram.

Only two quasars among $\sim$1,400 sources with $is$ and $iz$ photometry were newly identified as high redshift quasars
in the $r-is-iz$ color-color diagram, and none of our candidates were  discovered in the $is-iz-J$ color-color diagram. The other quasars were selected as candidates using the $WISE$ photometry or the color cuts from \citet{mcgr13}. 
The expected numbers of quasars for each selection method from 3,400 deg$^2$ are $24.4^{+67.7}_{-17.9}$ for 4.60 $\leq$ z $\leq$ 5.40 and $5.6^{+15.4}_{-4.1}$ for 5.50 $\leq$ z $\leq$ 6.05 (Section \ref{effi_numden}).
For 4.60 $\leq$ z $\leq$ 5.40, the number of quasars that we found is 6; including 14 previously discovered quasars in the literature, the total number of quasars is 20, which is in agreement with the expected number.
The selection for 5.50 $\leq$ z $\leq$ 6.05 identified two quasars 
that were published in previous studies, while we were unable to discover new quasars so far
(see Section \ref{effi_numden}), this number is also as expected.

\subsection{Spectroscopic Properties}\label{prop_phys}

\begin{figure}
\centering
\epsscale{1.2}
\plotone{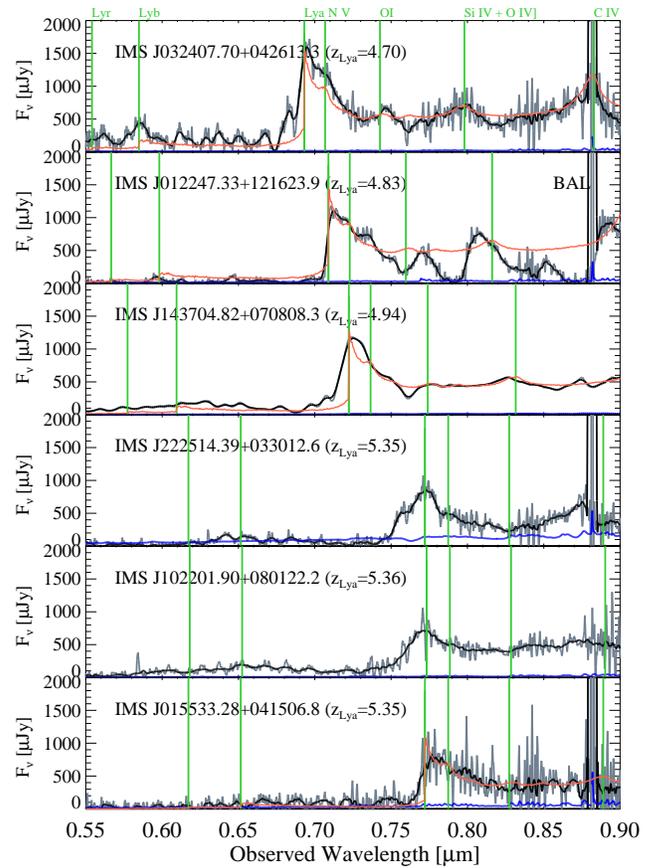}
\caption{Optical spectra of the 6 quasars at 4.7 $<$ z $<$ 5.4 from KPNO 4-m telescope/RCSPL and NTT/EFOCS2.
The gray lines are the original (oversampled) spectra and the black lines are spectra smoothed to their respective instrumental resolution. The blue lines denote the errors of each spectrum.
The orange line represents the redshifted composite spectrum of SDSS quasars \citep{vand01} 
including the IGM attenuation \citep{mada96}, 
which is fit to the observed spectrum. 
The green lines indicate general quasar emission lines.   
\label{f_prop_04}}
\end{figure}

First, we present the optical spectra of the 6 quasars at 4.7 $\leq$ z $\leq$ 5.4 in  
Figure \ref{f_prop_04}.
We plotted spectra smoothed to the resolution of each instrument (black lines)
together with the original spectra (gray lines).
The blue lines denote the errors of the spectra.

\begin{figure}
\centering
\includegraphics[scale=0.56]{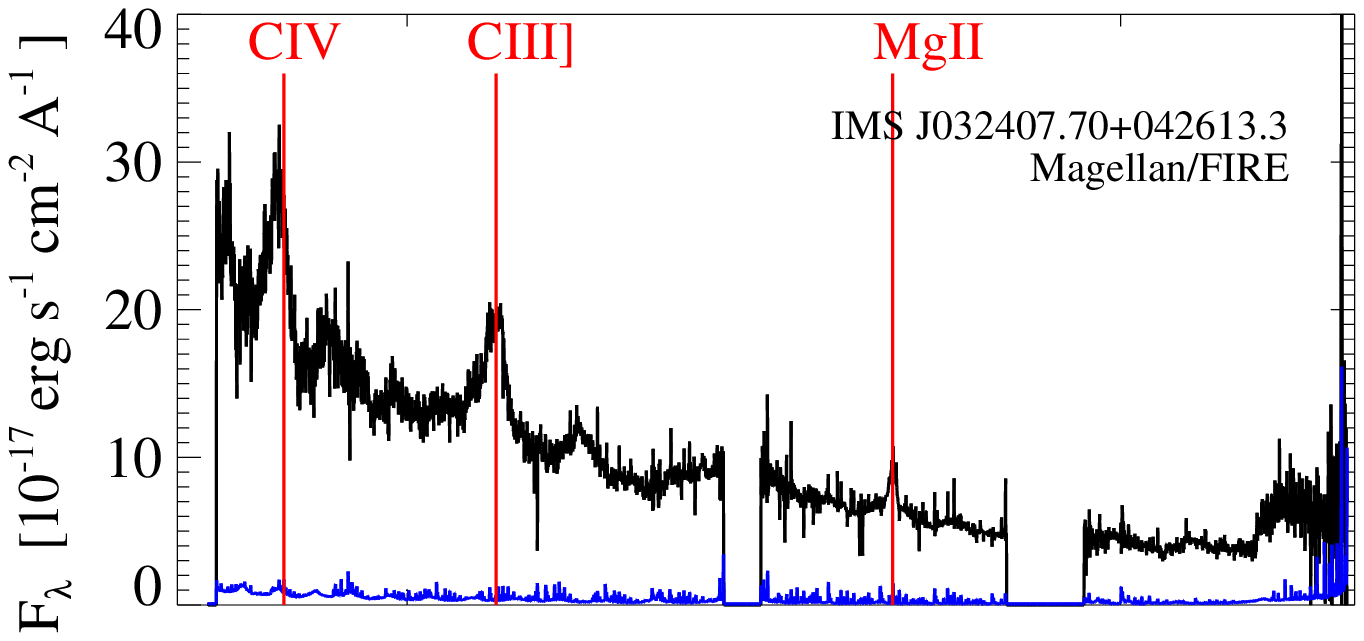}
\includegraphics[scale=0.56]{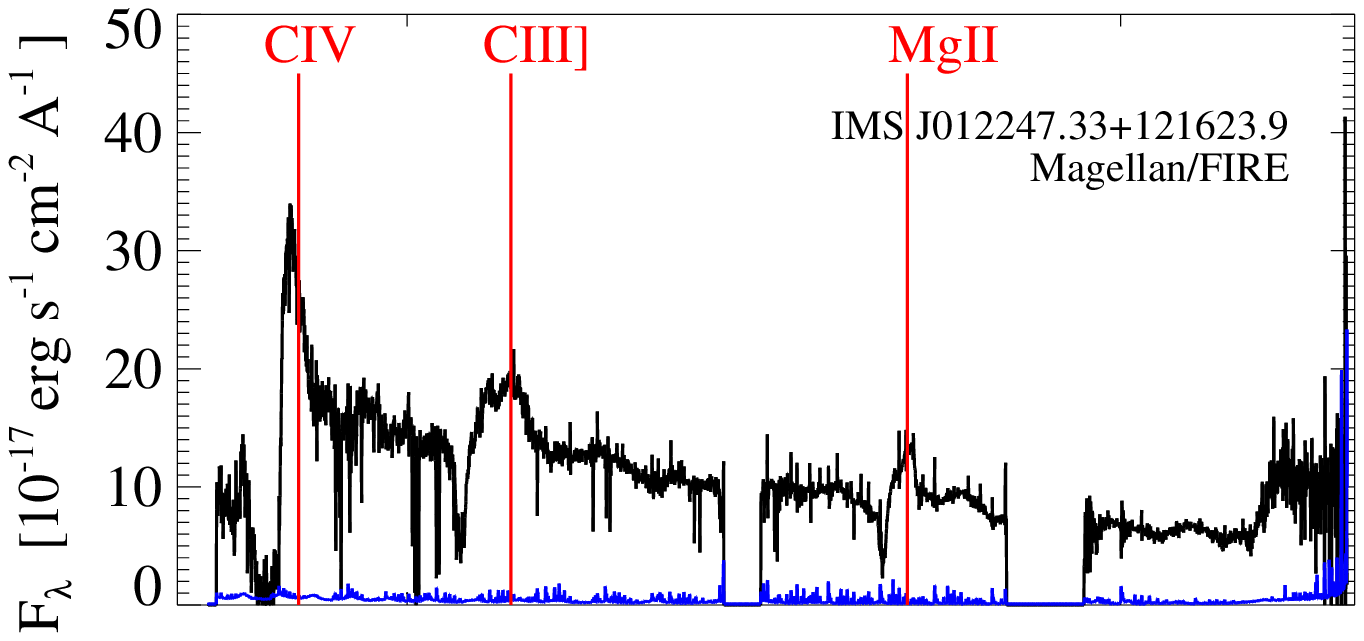}
\includegraphics[scale=0.56]{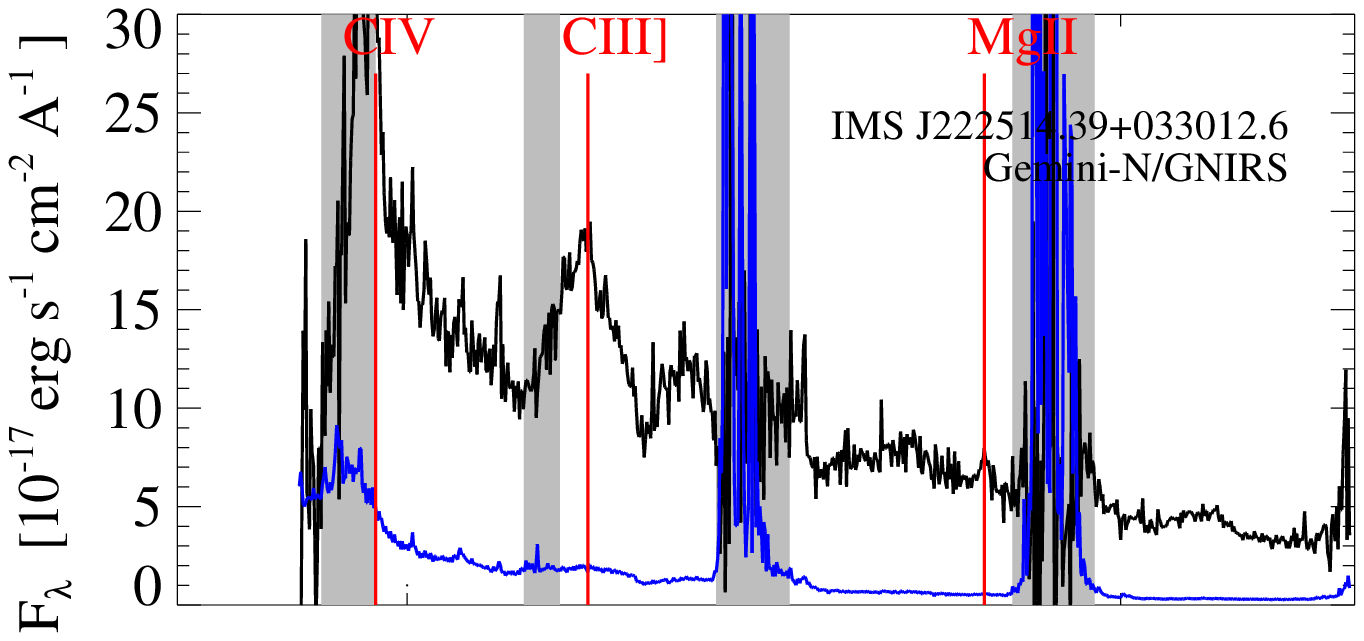}
\includegraphics[scale=0.56]{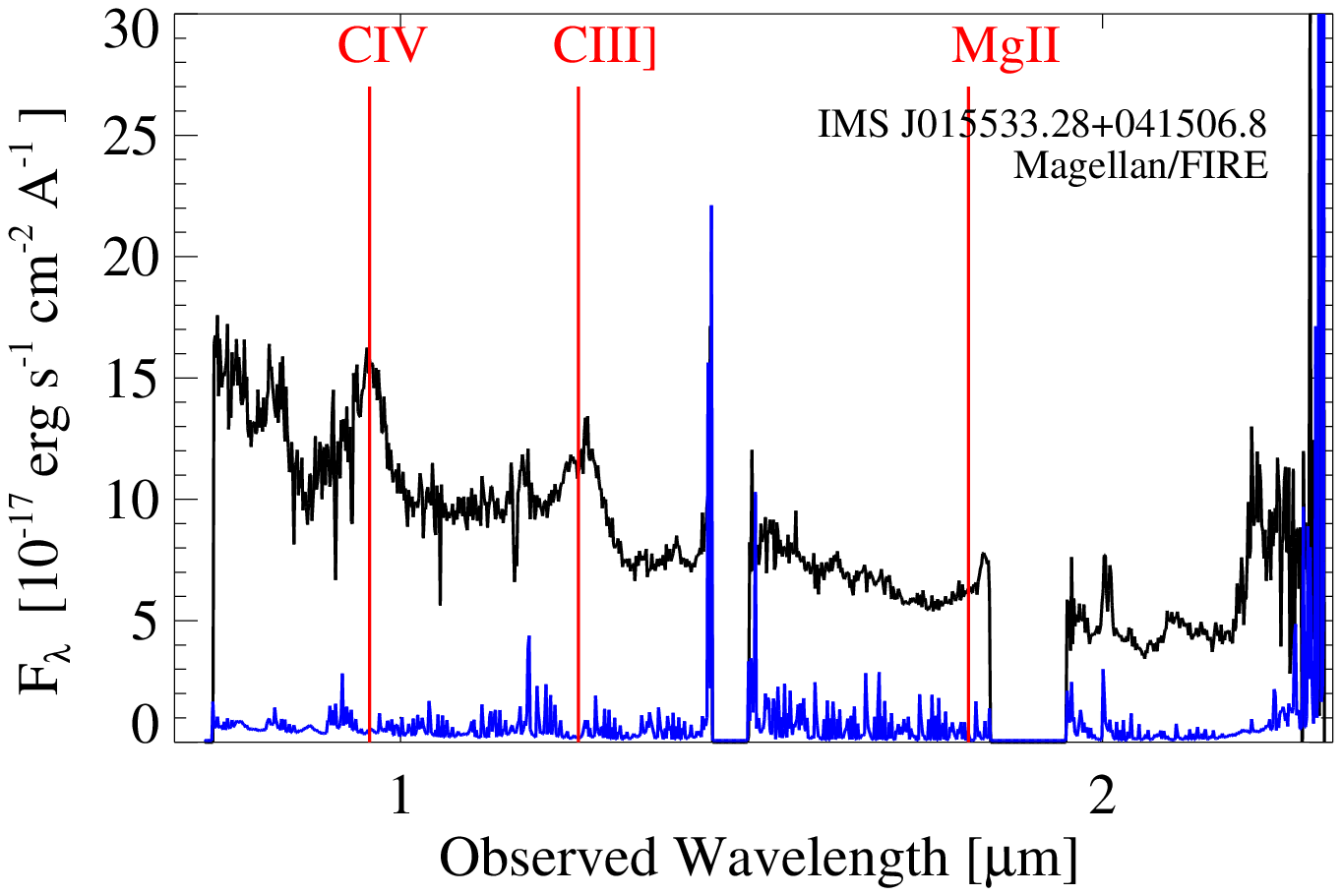}
\caption{
NIR spectra of IMS quasars 
from Magellan/FIRE and Gemini-N/GNIRS, smoothed to the instrumental resolution. 
The blue lines denote the errors of the spectra and the red vertical lines indicate 
the locations of emission lines at the redshift determined from optical spectra.
For the spectrum from Gemini-N/GNIRS, the gray bars show regions of 
strong atmospheric absorption.
\label{f_prop_05}}
\end{figure}

Second, we present NIR spectra of four objects, 
IMS J0324+0426, IMS J0122+1216, IMS J2225+0330 and IMS J0155+0415
in Figure \ref{f_prop_05}.
The reduced spectra were binned to the spectral resolution of each instrument using the median statistics. 
Errors of the smoothed spectra were calculated from the errors of the original spectra via standard error propagation. 
For the spectrum from Gemini-N/GNIRS, the gray bars show regions of 
strong atmospheric absorption, 
where the spectra shows low S/N. 

We find diverse Ly$\alpha$ shapes for the six quasars. IMS J0324+0426, IMS J0122+1216, and IMS J1437+0708 show strong Ly$\alpha$ emission, while the other three show smoother shapes. 
These weak Ly$\alpha$ lines are fairly common at high redshift. 
\citet{jian09} and \citet{bana14} show that a significant fraction of quasars at high redshift have weak Ly$\alpha$
\citep[e.g., 25\% of z $\sim$ 6 quasars discovered by][]{bana14}.
Most of the emission lines with the exception of Ly$\alpha$ are difficult to verify due to the imperfect 
sky line subtraction and low QE of the detector at wavelengths longer than 0.8 $\mu$m. 
IMS J0122+1216 shows significant deep absorption 
features and we classify it as 
a broad absorption line (BAL) quasar. 
This property can be noticed in its NIR spectrum more clearly.  

Table \ref{t_imsq} lists the redshifts and absolute magnitudes of the continua at rest-frame 1450$\AA$ ($M_{1450}$) of the quasars. 
The redshifts of IMS J2225+0330 and IMS J1022+0801 were measured from the Ly$\alpha$ emission lines by fitting Gaussian profiles. 
However other spectra show a sharp drop bluewards of Ly$\alpha$. 
In these cases, their redshifts were measured by fitting the spectra 
(the orange line in Figure \ref{f_prop_04})
from the redshifted and IGM-attenuated SDSS composite quasar template. 
The redshift errors estimated from these optical spectra contain the uncertainties from the spectral resolution of each instrument (typically $\sim$0.05),
because  one of the most dominant uncertainties of the redshift measurement is caused by the low spectral resolution. 
Also we list the redshifts estimated using the \civ\, or \mgii\, emission lines 
from the NIR spectra (see section \ref{bhmass}) in Table \ref{t_imsq}.
The redshift error estimated from the NIR spectra due to the spectral resolution 
is about 0.002 for Magellan/FIRE and about 0.007 for Gemini-N/GNIRS.
The redshifts estimated from the optical spectra and the NIR spectra show
discrepancies, and we believe that this is  
caused by the ambiguous Ly$\alpha$ shapes, 
which can be heavily affected by the Ly$\alpha$ forest and the blending with the \nv\, emission line.

\citet{rich09, rich15} provide photometric redshifts (z$_{\rm phot}$) for three out of the six quasars, IMS J0122+1216, IMS J1437+0708, and IMS J2225+0330.
Their estimate for IMS J0122+1216 
\citep[z$_{\rm phot}=5.455^{+0.135}_{-0.095}$ from][]{rich15}
does not agree with our spectroscopic redshift (z$_{\rm Ly\alpha}$ = 4.83), 
while IMS J1437+0708 
(z$_{\rm phot}=5.075^{+0.505}_{-0.455}$ from Richards et al. 2009 or $5.265^{+0.115}_{-0.505}$ from Richards et al. 2015)
and IMS J2225+0330 
\citep[z$_{\rm phot}=5.415^{+0.285}_{-0.395}$ from][]{rich15} 
are in agreement with our estimates (z$_{\rm Ly\alpha}$ = 4.94 and 5.35, respectively).
The discrepancy between z$_{\rm phot}$ and z$_{\rm spec}$ for IMS J0122+1216 is 
likely because the object is a BAL quasar. 

We calculated the $M_{1450}$ values using the average flux at 1440$\AA$ -- 1460$\AA$ from the optical spectra in Table \ref{t_imsq}. 
The uncertainties were estimated from the rms continuum flux density.
For z = 5.0 quasars, the observed wavelength of the rest-frame 1450$\AA$ is located at 8700$\AA$, where the sky emission lines are significant. Due to the difficulty of subtracting the sky from the relatively low S/N spectra, these values are crude and the actual magnitude uncertainties could be higher than our error estimates.
Our IMS quasars are within the $M_{1450}$ range of $-27.4$ -- $-26.4$.

\subsection{Individual Properties of Quasars}\label{prop_indi}

\textbf{IMS J0324+0426 (z$_{\rm Ly\alpha}$=4.70, z$_{\rm CIV}$=4.68, z$_{\rm MgII}$=4.73):}
This quasar has a strong Ly$\alpha$ emission line. It also shows relatively strong Lyman $\beta$ (Ly$\beta$), \oi\,, \siiv+\oiv, and \civ\, emission lines, and a weak \nv\, emission line. 
In the NIR spectrum, \civ\,, \ciii\,, and \mgii\, emission lines are prominent. 
\citet{wang16} reported z=4.72.

\textbf{IMS J0122+1216 (z$_{\rm Ly\alpha}$=4.83, z$_{\rm CIV}$=4.81):}
We classify this as a BAL quasar because of deep absorption features bluewards of Ly$\alpha$, \oi\,, \siiv+\oiv, and \civ\, lines.
It has a strong Ly$\alpha$ emission line, and a weak Ly$\beta$ emission line. 
We are not able to identify other emission lines due to these deep absorptions.   
The NIR spectrum has strong \civ\,, \ciii\,, and \mgii\, emission lines. 
The left side (shorter wavelengths) of these lines are severely absorbed. 
\citet{yi15} analyzed this quasar and derived a redshift of z=4.76
while \citet{wang16} reported z=4.79.

\textbf{IMS J1437+0708 (z$_{\rm Ly\alpha}$=4.94):}
Its spectrum was obtained from NTT/EFOSC2 with R $\sim$ 130 and  
it has the highest S/N ratio among the optical spectra. 
However it does not show any prominent emission lines except the Ly$\alpha$.
\citet{wang16} reported z=4.93.

\textbf{IMS J2225+0330 (z$_{\rm Ly\alpha}$=5.35 and z$_{\rm MgII}$=5.26):}
This source was observed by two telescopes, the KPNO 4-m telescope and NTT, and 
the two spectra were combined in average. 
It has a smooth Ly$\alpha$ emission line and does not show any other emission lines.
In the NIR spectrum, the \civ\,, \ciii\,, and \mgii\, emission lines are strong  
but the \civ\, emission line has a rough shape due to the strong atmospheric absorption.
\citet{wang16} reported z=5.24.
      
\textbf{IMS J1022+0801 (z$_{\rm Ly\alpha}$=5.36):}
This quasar has the weakest Ly$\alpha$ emission line among the six observed quasars.  
No other emission lines are visible due to low S/N. This quasar was recently discovered independently by \citet{yang17}, reporting the spectroscopic redshift of $z=5.30$.

\textbf{IMS J0155+0415 (z$_{\rm Ly\alpha}$=5.35, z$_{\rm CIV}$=5.27):}
The optical spectrum shows a weak Ly$\alpha$ emission line and other emission lines
are not detected. 
In the NIR spectrum, it has prominent \siiv+\oiv, \civ\,, and \ciii\, emission lines. 
The  \mgii\, emission line is hidden due to telluric absorption.
\citet{wang16} reported z=5.37.

\section{Selection Completeness}\label{effi}

To calculate the expected number of quasars for each selection method, we derived the quasar selection completeness, which can be affected by various effects. 
The completeness from color selection is defined as the fraction of quasars inside specific color cuts among all quasars within specific redshift and magnitude bins.
First, applying various quasar templates, we calculated the completeness using the fraction of quasars that fall within each selection box, as a function of redshift and $M_{1450}$ (Section \ref{effi_color}).
Then we apply this completeness to our quasar surveys and predict the expected quasar number of each selection method in Section \ref{effi_numden}. 

\subsection{Completeness from Color Cuts} \label{effi_color}

To measure the fraction that a quasar with a given redshift, $M_{1450}$, and intrinsic SED meets our selection criteria, 
we follow approaches from previous studies 
\citep[e.g.,][]{will05, vene13}. The composite quasar template from \citet{vand01} is redshifted to various values, assuming that the spectral properties of quasars do not evolve significantly with redshift 
\citep[e.g.,][]{kuhn01, fan04, jun15}, except wavelengths blueward of the Ly$\alpha$ line. 
Fluxes in these shorter wavelengths are absorbed by neutral hydrogen (\hi) in the IGM, and the absorption becomes stronger toward higher redshift because the fraction of \hi~ increases with redshift 
\citep[Gunn-Peterson effect;][]{gunn65}. We applied this attenuation effect to our redshifted spectra using the IGM attenuation model of \citet{mada96}. We redshifted the spectrum to 4 $\leq$ z $\leq$ 8 with steps of $\Delta$z = 0.05 and adopted $M_{1450}$ in the range  
$-30 < M_{1450} < -20$ with steps of $\Delta M_{1450}=0.5$. Then we calculated model magnitudes for each band.

The most important factor in the observed color distribution 
is the continuum slope of quasars. 
We considered 13 cases of models for each redshifted spectrum with continuum slopes of $-1.3 \leq \alpha_{\nu} \leq -0.1$ (where $F(\nu) \propto \nu ^ {\alpha_{\nu}}$) with steps of $\Delta \alpha_\nu = 0.1$. 
This range was derived based on the range of $\alpha_\nu$ values from the SDSS DR12 quasar catalog \citep{pari16} 
that includes about 230,000 quasars with a mean value of $\alpha_{\nu}=-0.7$ and  a 1$\sigma$ dispersion of 0.6 (68.3\% confidence level). 
\citet{dero14} analyzed a sample of four quasars at z $>$ 6.5 and three of these four quasars (75\%) fall in this $\alpha_\nu$ range. 
We also considered variable rest-frame equivalent widths (EW$_{0}$) of the Ly$\alpha$ emission line: 8 cases of 50 $\leq$  EW$_{0}$ $\leq$ 85 with steps of $\Delta$EW$_{0}$ = 5 \citep{fan01}. 
In total, we generate a database of 104 model quasars 
of which the continuum slopes and Ly$\alpha$ EWs are uniformly sampled within given ranges
and calculate the average selected fraction as a function of redshift and $M_{1450}$.

\begin{figure}
\centering
\includegraphics[scale=0.45]{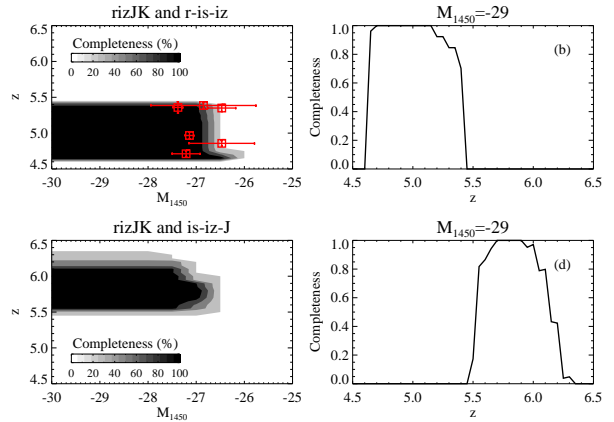}
\caption{ 
(a): Completeness as a function of redshift and $M_{1450}$ for $r-i-z-J-K$ and $r-is-iz$ selection.
The red boxes indicate the redshifts and $M_{1450}$ of our six new quasars.
(b): Completeness as a function of redshift from (a) when $M_{1450} = -29$. 
(c): Completeness for $r-i-z-J-K$ and $is-iz-J$ selection.
(d): Completeness from (c) when $M_{1450} = -29$. 
The colors of the contours indicate 0\% and 100\% completeness for white and black, respectively.
\label{f_effi_01}}
\end{figure}

Figure \ref{f_effi_01}a shows the completeness distribution as a function of redshift and $M_{1450}$, for the selection using the $r-i-z-J-K$ 
and $r-is-iz$ color-color diagrams (selection method A), and Figure \ref{f_effi_01}c shows the completeness distribution when using $r-i-z-J-K$ and $is-iz-J$ color-color diagrams (selection method B).
In Figures \ref{f_effi_01}b and \ref{f_effi_01}d,
we plot the completeness as a function of redshift for the two methods, 
for the case of $M_{1450} = -29$. The completeness in Figure \ref{f_effi_01}b rises steeply from 0\% to 100\% between z = 4.60 and z = 4.70, remains at 100\% up to z = 5.15, and drops below 80\% for z $>$ 5.35. In the case of Figure \ref{f_effi_01}d, the slopes of the completeness distribution at the borderline redshift values are more gradual than those in Figure \ref{f_effi_01}b. The redshift ranges of the completeness greater than 80\% are 4.60 $\leq$ z $\leq$ 5.40 for method A and 5.50 $\leq$ z $\leq$ 6.05 for method B, which represent  the expected redshift ranges of quasars selected from the two color-color diagrams. 
The completeness of both selection methods drop to below 50\% at $M_{1450} > -27.0$ when z = 4.90 and z = 5.80, 
where the $M_{1450}$ limit corresponds to our magnitude cut, $z < 19.5$ mag. We also plot the redshifts and $M_{1450}$ of our six newly discovered quasars
(Table \ref{t_imsq}) with red boxes in Figure \ref{f_effi_01}a.

\subsection{Expected Quasar Number from Our Surveys}\label{effi_numden}

\begin{deluxetable*}{c|c|c|c|cc|c}
\tabletypesize{\scriptsize}
\tablecaption{
Expected number of quasars from our survey
 \label{t_surv}}
\tablewidth{0pt}
\tablehead{
Selection Method	&Area (deg$^2$)	&Redshift Range	&$M_{1450}$  Limit	&\multicolumn{2}{c|}{Expected Number} &Selected Number\\
(1)			&(2)		&(3)		&(4)			&(5)\tablenotemark{a} & (6)\tablenotemark{b}& (7)
}
\startdata
$r-is-iz$&	3,400		&4.60 --  5.40	&$-27.0$		&    $24.4^{+67.7}_{-17.9}$ & $47.3^{131.2}_{-34.7}$&20\\ 
$is-iz-J$&	3,400		&5.50 -- 6.05	&$-27.0$		&    $5.8^{+15.9}_{-4.3}$ & $6.9^{+19.0}_{-5.1}$&2 
\enddata
\tablenotetext{a}{
For $k=-0.47$
}
\tablenotetext{b}{
For $k=-0.71$
}

\end{deluxetable*}

We calculated the expected number of quasars from our survey by extrapolating 
the luminosity function of z $\sim$ 6 quasars from \citet{will10139}.
We considered the $10^{kz}$ factor that accounts for the decline in number density 
as a function of redshift. 
We adopted two values of $k$: $k=-0.47$ from \citet{will10139}  and $k=-0.71$ from \citet{mcgr13}. 
Then, we extrapolated the luminosity function of z $\sim$ 6 to our redshift range, and derived the expected number of quasars from our survey. 
Table \ref{t_surv} shows our quasar selection with different selection methods (column 1), survey area (column 2), redshift range (column 3),
and $M_{1450}$ limit (column 4).
The expected number of quasars for each quasar selection are listed in columns 5 and 6 for the case of $k=-0.47$ and $k=-0.71$, respectively, 
with the 1$\sigma$ errors caused by the uncertainties in break magnitude $M_{1450} ^*$ and bright end slope $\beta$ from \citet{will10139}. 
We only considered the completeness from our color cuts, and assumed that the efficiency of each selection in its redshift range (column 3)  and the $M_{1450}$ limit (column 5) is 100\%.

Our quasar survey discovered 20 quasars including 6 new quasars at 4.60 $\leq$ z $\leq$ 5.40. This number is consistent with that from the luminosity function at 4.60 $\leq$ z $\leq$ 5.40. However we could not find any new quasars at 5.50 $<$ z $<$ 6.05, 
except two previously discovered quasars.
We believe that the absence of any new quasars at 5.50 $<$ z $<$ 6.05 is due to the lack of $WISE$ photometry (they are fainter than quasars at 4.60 $\leq$ z $\leq$ 5.40), 
resulting in a lower priority for the CQUEAN imaging. We expect to uncover more promising candidates as we build up the CQUEAN follow-up imaging sample.

\section{Physical Properties of Quasars}\label{bhmass}

In this section, we present the physical properties of four IMS quasars, 
IMS J0324+0426, IMS J0122+1216, IMS J2225+0330 and IMS J0155+0415,  
based on the data obtained with optical and NIR spectroscopy.
In our NIR spectra, we identified both the \civ\, and \mgii\, lines for IMS J0324+0426 and IMS J0122+1216, 
only the \mgii\, line for IMS J2225+0330, 
and only the \civ\, line for IMS J0155+0415.
After modeling the continuum and emission lines of \civ\, and \mgii\,, we estimated 
continuum slopes $\alpha_{\nu}$ 
(where $\alpha_{\nu}$  is  for $ F(\nu) \propto \nu ^{ \alpha_{\nu} }$), 
line widths (full width at half maximum; FWHM), 
continuum luminosities at the rest-frame wavelengths of 1350$\AA$ and 3000$\AA$
($\lambda L_{\lambda}(1350)$ and $\lambda L_{\lambda}(3000)$)
for each emission line (Section \ref{bhmass_anal}). 
From these measurements, we calculated the black hole mass ($M_{\rm BH}$) from the \civ\, emission line ($M_{\rm BH,CIV}$) or from the \mgii\, emission line ($M_{\rm BH,MgII}$) through different relations from \citet{mclu04}, \citet{vest06}, and \citet{jun15} (Section \ref{bhmass_result}). 
For the virial factor in these black hole mass estimators, we adopted $f=5.1\pm1.3$ from \citet{woo13}. 
In Section \ref{bhmass_eddra}, we compare the  Eddington ratios of our quasars 
to lower redshift quasars.

\subsection{Analysis of NIR Spectra}\label{bhmass_anal}

We modeled the quasar NIR continuum
assuming two components, a power law component and a component that 
describes the pseudo-continuum due to the blended forest of
 \feii\, emission lines as given below:
\begin{equation}
F(\lambda) = a \times \lambda ^{ \alpha_{\lambda} } +  b \times {\rm FeII} (\lambda, v) 
\end{equation}
where $\alpha_{\lambda}$ is the continuum slope
(in this case, $\alpha_{\nu}  =  -\alpha_{\lambda} - 2 $ for $ F(\lambda) \propto \lambda ^{ \alpha_{\lambda} }$), and 
$v$ and $b$ are the width and strength of \feii\,\, templates.
We used two \feii\, templates from \citet{vest01} and \citet{tsuz06}. 
A scaled and broadened \feii\, template was used for modeling the \feii\, emissions from our spectra. 
In the case of the \civ\, emission line, only \citet{vest01} provide the \feii\, template in this wavelength range. We modeled the two components simultaneously. 

The quality of the continuum subtraction depends on the determination of the continuum fitting ranges. We selected narrow fitting windows which minimize the contributions from other components.
Since the qualities of the \civ\, emission line in the IMS J0324+0426 spectrum 
and the \mgii\, emission line in the IMS J2225+0330 spectrum are 
not sufficient to constrain the \feii\, emissions, we failed to find the \feii\, component.
Since IMS J0122+1216 shows significant broad absorption features bluewards of the \civ\, and \mgii\, emission lines, 
we narrowed the fitting window ranges to exclude the absorption part.

Since most of the uncertainties in the continuum slope or the line width result from the fitting range of the continuum modeling,  
we adopted 36 different fitting ranges within the given wavelength windows 
and performed model fitting for each different sub-wavelength range 
to calculate the uncertainties. 
Since we cannot vary the continuum fitting range of \civ\, of IMS J0122+1216,  
we set the uncertainty of this line width as 5\% of the line width instead of the uncertainty derived from the various continuum ranges.
This fraction is identical to the ratio of line widths and their uncertainties, 
for all other lines.

\begin{figure*}
\centering
\includegraphics[scale=0.65]{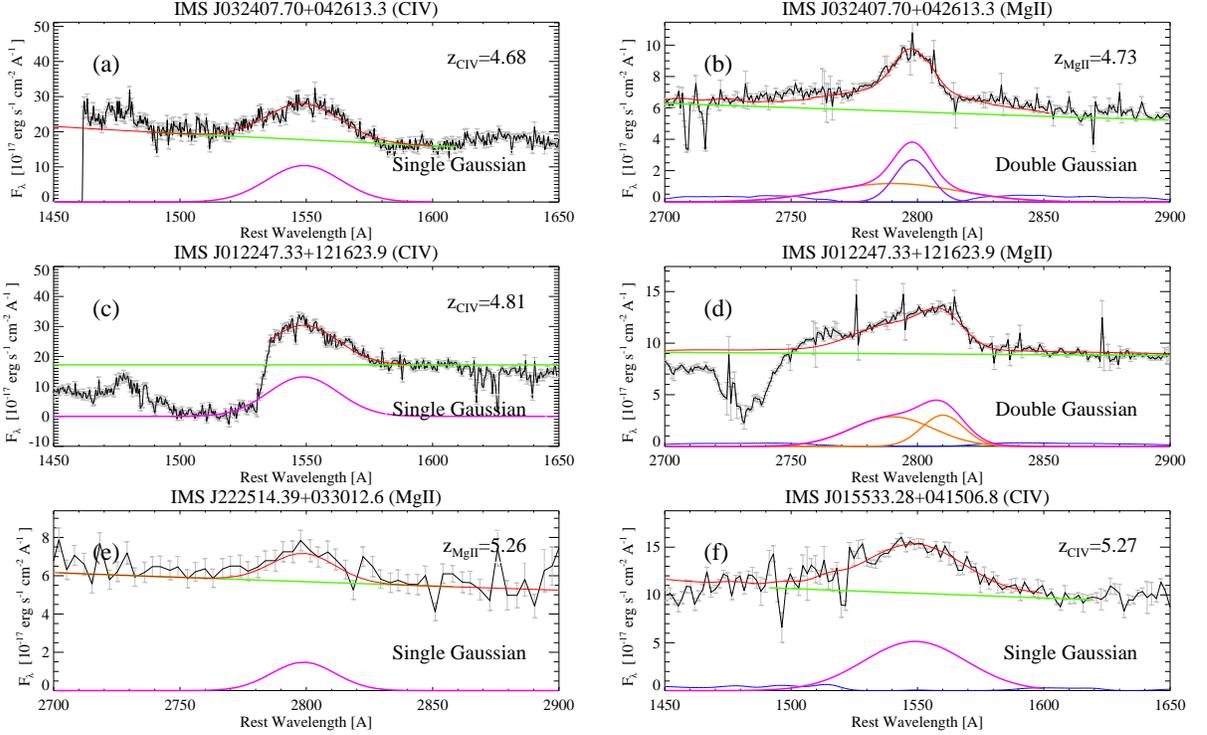}
\caption{ 
The best-fit continuum and emission line modeling for the 
\civ\, and \mgii\, emission lines of IMS quasars.  
In each panel, the spectrum (black) with errors (gray) is overplotted 
with the best-fit model (red), which consists of the power-law component (green), 
the best-fit \feii\, template (blue; we used the \feii\, template from  Vestergaard \& Wilkes 2001  as an example), and each emission line (magenta).  
Estimated redshift from each emission line is denoted except for the \mgii\, emission line of IMS J0122+1216.
(a): \civ\, emission line of IMS J0324+0426.
We cannot find a solution for the \feii\, template fitting. 
(b): \mgii\, emission line of IMS J0324+0426.
We used a double-Gaussian model for the line fitting (violet and orange lines)
and one of the double-Gaussian components is a narrow line (violet line).
(c): \civ\, emission line of IMS J0122+1216.
(d): \mgii\, emission line of IMS J0122+1216.
Two Gaussian components (two orange lines) are used to fit the line shape. 
(e):  \mgii\, emission line of IMS J2225+0330.
We cannot find a solution for the \feii\, template fitting. 
(f): \civ\, emission line of IMS J2225+0330.
\label{f_bhmass}}
\end{figure*}

After subtracting the best-fit continuum from each spectrum, 
we fit the \civ\, and \mgii\, emission lines.
We used single and double-Gaussian profiles considering the presence of asymmetric profiles characterized by red or blue wings. 
For the fitting ranges, we set 
1500$\AA$ -- 1600$\AA$ for the \civ\, line 
and 2700$\AA$ -- 2900$\AA$ for the \mgii\, line,  
except for the \civ\, of IMS J0122+1216, which is affected by broad absorption. 
In this case, we set the fitting range to 1530$\AA$ -- 1590$\AA$.
The \mgii\, lines of IMS J0324+0426 and IMS J0122+1216 are well fit by double-Gaussian profiles due to their asymmetric shapes, whereas the other lines can be fit using a single-Gaussian profile. 
One of the double-Gaussian components of IMS J0324+0426 is a narrow line (violet line in Figure \ref{f_bhmass}b) with FWHM = 800$\pm$40 km s$^{-1}$.
To obtain the line width FWHM, the measured FWHM$_{\textrm{obs}}$ was corrected for the instrumental resolution FWHM$_{\textrm{ins}}$:
$\textrm{FWHM} = \sqrt{(\textrm{FWHM}_{\textrm{obs}})^2 - (\textrm{FWHM}_{\textrm{ins}})^2  }$.

We used an IDL procedure, \texttt{mpfit.pro} to find the best-fit models to the observed spectra that uses the $\chi^2$ minimization method for both the continuum and the emission line. 
We included 1$\sigma$ errors of the spectra for each fitting. 
From the best-fit model, 
we obtained the best-fit estimates for each parameter, such as the power law slope and the line width. 
The uncertainties for each parameter were calculated as follows.
The error for each parameter is dominated by the scatter of the various best-fits when 
altering the fitting range for the continuum. 
We compared the best-fit parameters for each trial 
and we set the average and standard deviation of the values as the best-fit parameter and its error.  

\subsection{Ultraviolet Luminosity and $M_{\rm BH}$} \label{bhmass_result}

\begin{deluxetable*}{crrcccccccccccc}
\tabletypesize{\scriptsize}
\tablecaption{Power-law slopes, line widths, and continuum luminosities estimated from the NIR spectra\label{t_pr01}}
\tablewidth{0pt}
\tablehead{
Name&	$\alpha_{\nu,\rm CIV}$&	$\alpha_{\nu,\rm MgII}$&	FWHM$_{\rm CIV}$&	FWHM$_{\rm MgII}$&	$\lambda L_{\lambda}(1350)$&	$\lambda L_{\lambda}(3000)$ \\
&						&					&		(km s$^{-1}$)		&		(km s$^{-1}$)		&		($10^{46}$ erg s$^{-1}$)		&	($10^{46}$ erg s$^{-1}$)		
}
\startdata

IMS J0324+0426&$ 1.34\pm0.60$&$-0.42\pm0.78$&$6070\pm300$&$2660\pm280$&$  6.93\pm 2.22$&$  3.69\pm 0.35$\\
IMS J0122+1216&\nodata&$-1.57\pm0.31$&$6240\pm310$&$4210\pm160$&$  5.91\pm 0.08$&$  6.11\pm 0.64$\\
IMS J2225+0330&\nodata&$0.49\pm0.38$&\nodata&$2750\pm490$&\nodata&$4.08\pm0.22$\\
IMS J0155+0415&$-0.71\pm0.85$&\nodata&$8140\pm800$&\nodata&$  6.44\pm 0.48$&\nodata
\enddata
\end{deluxetable*}

In Figure \ref{f_bhmass},  
the best-fit continuum and emission line models are shown for each emission line. 
In Table \ref{t_pr01} we list the best-fit estimates of the power law slope ($\alpha_{\nu, \rm CIV}$ and $\alpha_{\nu, \rm MgII}$) and the line width (FWHM$_{\rm CIV}$ and FWHM$_{\rm MgII}$)  and their errors for each emission line.
There is no significant difference in the derived power law slope and line width parameters 
when using different \feii\, templates from \citet{vest01} and \citet{tsuz06}.
Note that the IMS 2225+0330 spectrum has low S/N and the uncertainty of the line width estimated using the method in Section \ref{bhmass_anal} is underestimated. 
The 1$\sigma$ error from the Gaussian fitting is about 15\%.

The power law slopes of quasars vary significantly between sources.
For example, \citet{davi07} found $-1.5 <$ $\alpha_{\nu}$ $<$ 0.5
for quasars at 0.76 $<$ z $<$ 1.26 and 1.67 $<$ z $<$ 2.07. 
At high redshift, 
quasars at 4 $<$ z $<$ 6.5 from \citet{dero11} showed $-4 <$ $\alpha_{\nu}$ $<$ 0.7, 
and quasars at z $>$ 6.5 from \citet{dero14} showed $-0.67 <$ $\alpha_{\nu}$ $<$ 0.56.
The slope coefficients from our results are in agreement with
these values at high redshift.

The $\lambda L_{\lambda}(1350)$ and $\lambda L_{\lambda}(3000)$ in Table \ref{t_pr01} are also calculated 
from the optical and NIR spectra. 
For IMS J0324+0426, we used the optical and NIR spectra for the $\lambda L_{\lambda}(1350)$ and $\lambda L_{\lambda}(3000)$, respectively.  
The $\lambda L_{\lambda}(1350)$  of IMS J0155+0415 and the $\lambda L_{\lambda}(3000)$ of IMS J0122+1216 were estimated from their NIR spectra.
Since the continuum spectra near the 3000$\AA$ of IMS J2225+0330
show low S/N due to the strong atmospheric absorption,
we used fit spectra using 
the redshifted SDSS composite quasar template.
In the case of the $\lambda L_{\lambda}(1350)$ of IMS J0122+1216, the continuum near 1350$\AA$ shows deep drops in its optical spectrum. 
Therefore, we used the fit spectrum when we estimated the redshift in Section \ref{prop_phys}.
The $\lambda L_{\lambda}(1350)$ and $\lambda L_{\lambda}(3000)$ were calculated from the average flux in the 1340$\AA$ -- 1360$\AA$ and 2950$\AA$ -- 3050$\AA$, respectively. The uncertainty in the continuum luminosity was estimated from the scatter on the continuum flux in each window.

\begin{deluxetable*}{ccccccccccccccc}
\tabletypesize{\tiny}
\tablecaption{$M_{\rm BH}$, $L_{\rm Bol}$,  $L_{\rm Edd}$, and Eddington ratios
 \label{t_pr02}}
\tablewidth{0pt}
\tablehead{
Name	&$M_{\rm BH,CIV}$&$M_{\rm BH,MgII}$		&$L_{\rm Bol}(1350)$   &$L_{\rm Bol}(3000)$&$L_{\rm Edd}(\rm CIV)$&$L_{\rm Edd}(\rm MgII)$ &Edd. ratio& Edd. ratio\\
		&($10^9$ \Msolar)&($10^9$ \Msolar)  	&($10^{47}$ erg/s)	&($10^{47}$ erg/s)    &($10^{47}$ erg/s)&		($10^{47}$ erg/s)	& (1350, CIV) & (3000, MgII)
}
\startdata
IMS J0324+0426& 7.60$\pm$ 1.55&1.17$\pm$0.31&2.6$\pm$ 0.8& 1.9$\pm$ 0.2&  9.6$\pm$2.0&   1.5$\pm$ 0.4&0.28$\pm$0.16&1.29$\pm$0.60\\
IMS J0122+1216& 7.38$\pm$0.78& 4.76$\pm$0.52&2.3$\pm$ 0.1& 3.1$\pm$ 0.3&  9.3$\pm$ 1.0& 6.0$\pm$ 0.7&0.24$\pm$0.04&0.53$\pm$0.16\\
IMS J2225+0330&\nodata&1.35$\pm$0.59&\nodata&2.1$\pm$0.1&\nodata&1.7$\pm$0.7&\nodata&1.24$\pm$0.91\\
IMS J0155+0415&13.53$\pm$2.87&\nodata&2.5$\pm$ 0.2&\nodata&17.1$\pm$3.6&\nodata&0.14$\pm$0.05&\nodata
\enddata

\end{deluxetable*}

In Table \ref{t_pr02}, we list the virial black hole mass estimates obtained from \civ\, and \mgii\, emission lines
($M_{\rm BH,CIV}$ and $M_{\rm BH,MgII}$) 
using relations presented in \citet{jun15}.
The uncertainties of the masses propagate from the uncertainties of the FWHM and the continuum luminosity.
The Eddington luminosities ($L_{\rm Edd}$) estimated from the two mass estimators are listed in Table \ref{t_pr02}.
Comparing the two mass estimates ($M_{\rm BH, CIV}$ and $M_{\rm BH, MgII}$) for IMS J0324+0426 and IMS J0122+1216, $M_{\rm BH,CIV}$ is larger than $M_{\rm BH,MgII}$ by 0.8 dex 
and 0.2 dex, respectively. 
We note that $M_{\rm BH}$ values from \civ\, show a larger scatter with respect to those from \mgii\, or H$\beta$/H$\alpha$ \citep[e.g.,][]{jun15,karo15}. 
For example, the intrinsic scatters of the $M_{\rm BH, CIV}$ and $M_{\rm BH, MgII}$ from \citet{jun15} is 0.40 dex and 0.09 dex, respectively.
Therefore the large discrepancy between $M_{\rm BH, CIV}$ and $M_{\rm BH, MgII}$ can be understood as a result of the large scatter in $M_{\rm BH, CIV}$ estimators. Hence, we take the \mgii\, based values to be more reliable. 
The $M_{\rm BH}$ values are roughly consistent with each other, when using different estimates 
(e.g. \citet{mclu04} or \citet{vest06})
that use the same emission line, within the error bars and the intrinsic scatter in the $M_{\rm BH}$ estimators.

\subsection{Accretion Rate of Newly Discovered Quasars } \label{bhmass_eddra}

Bolometric luminosities ($L_{\rm Bol}$) and Eddington ratios are given in Table \ref{t_pr02}, where $L_{\rm Bol}$ are computed from $\lambda L_{\lambda}(1350)$ and $\lambda L_{\lambda}(3000)$ by multiplying 3.81 and 5.15, respectively \citep{shen08}. 

For IMS J0122+1216, the $L_{\rm Bol}$ values that are calculated from $\lambda L_{\lambda}(1350)$ and $\lambda L_{\lambda}(3000)$ do not agree with each other.
Since the  $\lambda L_{\lambda}(1350)$ is estimated from the best-fit model spectrum, 
we adopt $\lambda L_{\lambda}(3000)$ as more reliable.
In the case of IMS J0324+0426, 
the  $L_{\rm Bol}(1350)$ has a larger uncertainty due to significant contamination from sky emission lines.

The Eddington ratios from $M_{\rm BH,CIV}$ and $\lambda L_{\lambda}(1350)$ are smaller by a factor of a few than those using $M_{\rm BH,MgII}$ and $\lambda L_{\lambda}(3000)$. The discrepancy is most likely caused by the difference in the derived $M_{\rm BH}$ values. As we mentioned earlier, \civ-based $M_{\rm BH}$ values are in general more uncertain than \mgii-based values, and therefore we consider \mgii-based Eddington ratios to be more reliable. 

\begin{figure*}
\centering
\includegraphics[scale=0.94]{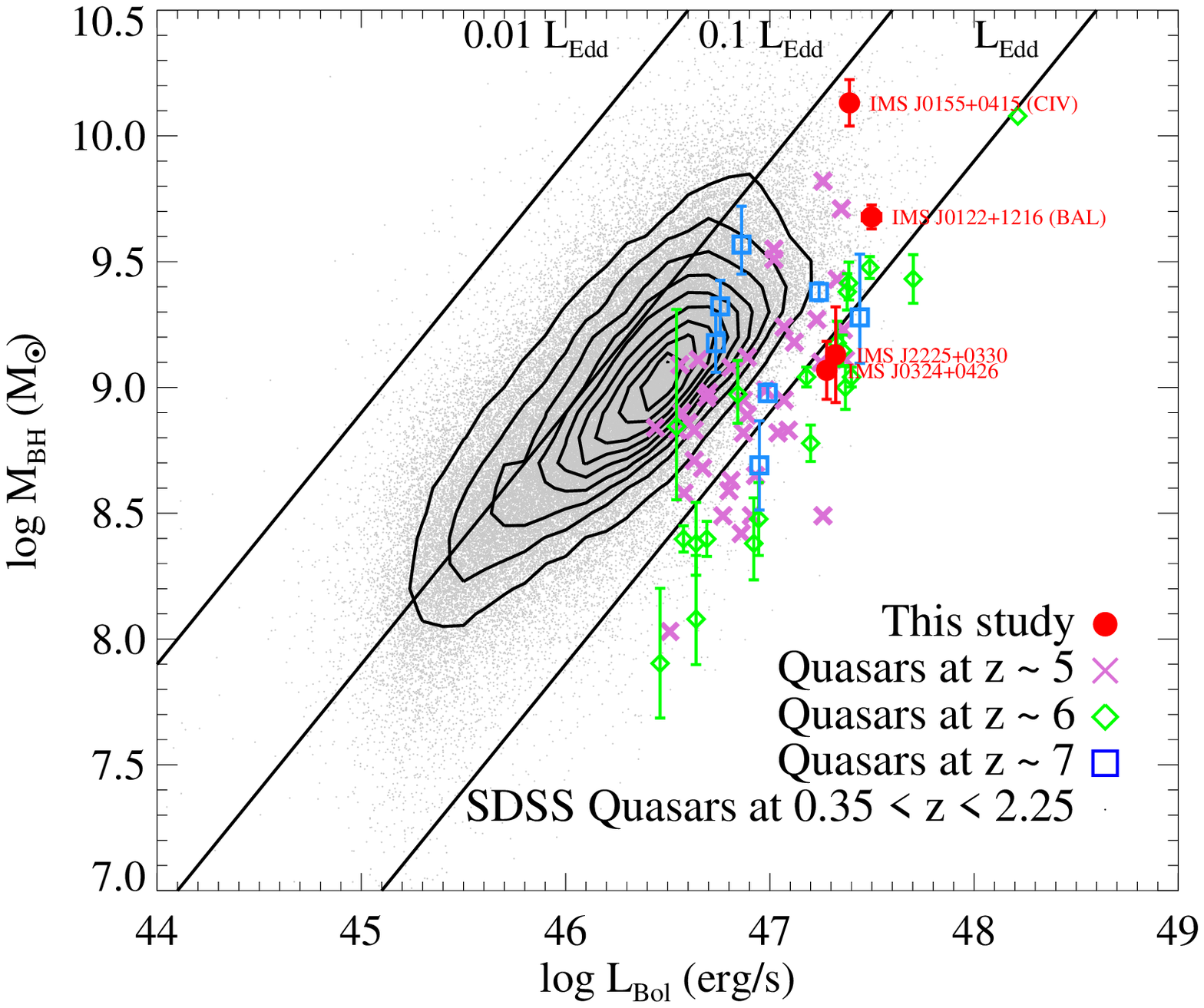}
\caption{ 
$M_{\rm BH}$ as a function of $L_{\rm Bol}$. 
Red filled circles are our sources, 
purple crosses are quasars at z $\sim$ 5 
\citep{trak11},
green diamonds are quasars at z $\sim$ 6 
\citep{jian07, kurk07, kurk09, wu15},
blue squares are quasars at z $\sim$ 7 
\citep{dero14, vene15}, and   
gray points and black contours are a subsample of SDSS quasars at 0.35 $<$ z $<$ 2.25 \citep{shen11}.
Lines of constant Eddington ratio for $L_{\rm Bol} /L_{\rm Edd}$ = 0.01, 0.1, and 1 
are plotted with black solid lines.
The names of the four newly discovered quasars are written next to the red filled circles. Names with 'BAL' and 'CIV' are for the less reliable $M_{\rm BH}$ values (either a BAL quasar, or $M_{\rm BH}$ estimated from the \civ\ line.)
\label{f_bhmass_11}}
\end{figure*}

Figure \ref{f_bhmass_11} shows $M_{\rm BH}$ as a function of $L_{\rm Bol}$. 
To compare our sources with low redshift quasars, we used the SDSS samples of quasars \citep{shen11}. Quasars with $M_{\rm BH,MgII}$ information were selected and they cover a redshift range of 0.35 $<$ z $<$ 2.25 (gray points and black contours). 
We also include 
quasars at 
z $\sim$ 5 
\citep[][purple crosses]{trak11},  
z $\sim$ 6 
\citep[][green empty diamonds]{jian07, kurk07, kurk09, wu15}, and 
z $\sim$ 7 
\citep[][blue empty squares]{dero14, vene15}.
All $M_{\rm BH}$ values are derived using \mgii\, estimators.
The Eddington ratios,  $L_{\rm Bol} /L_{\rm Edd}$ = 0.01, 0.1, and 1, 
are indicated with black solid lines.  
Our sources are plotted with the red filled circles
from $M_{\rm BH,MgII}$ and $L_{\rm Bol}(3000)$ except IMS J0155+0415. 
We can see that the high redshift sample occupies a region of the parameter space different from that of the low redshift sample with similar $L_{\rm Bol}$: the Eddington ratios of these high redshift quasars are significantly larger than those of the low redshift sample. 
In particular, our high redshift quasars have Eddington ratios around 1, 
suggesting that these quasars are growing vigorously. 
The Eddington ratio of IMS J0155+0415 is an exception, because it  
was estimated from $M_{\rm BH,CIV}$ and  $L_{\rm Bol}(1350)$, which are less reliable than $M_{\rm BH,MgII}$ and  $L_{\rm Bol}(3000)$, respectively.
\citet{will10140} show similar results that the luminosity-matched quasar samples at z = 2 and z = 6
have different Eddington ratio distributions. However, to compare the Eddington ratio distribution of low redshift
quasars to their high redshift counterparts, less luminous samples will be needed. 
Intrinsic Eddington ratios of normal high redshift quasars can be studied by discovering quasars from deeper surveys \citep[e.g.,][]{kash15, kim15} and Eddington ratio distributions at high redshift when less
luminous quasars are included can be different (e.g., Kim et al. in preparation).

\section{Summary} \label{summary}

We conducted a quasar survey at $5 \lesssim {\rm z} \lesssim 5.7$ 
using multi-wavelength data with new selection techniques.
First, candidates were selected from our $r-i-z-J-K$ color cuts, 
then we exploited the $WISE$ colors to narrow down the candidates. 
The candidates were also observed with the CQUEAN $is$ and $iz$ filters
that overcome the limitations of previous filter systems.  
We then carried out optical spectroscopic observations 
to confirm our high redshift quasar candidates and discovered six new quasars. 
Four of them  were observed by NIR spectroscopy to measure their physical properties
($M_{\rm BH}$, $L_{\rm Bol}$, $L_{\rm Edd}$, and Eddington ratio)
via spectral modeling of their continuum and emission lines. 
We compared Eddington ratios of our sources to those of low and high redshift quasars, 
and found that the Eddington ratio of our quasars at z $\sim$ 5 have values close to 1. 
These results, 
characterized by high luminosities ($M_{1450} < -27$ mag), larger black hole masses of $>10^9\Msolar$, and near-Eddington limit luminosities,
support the scenario of rapid growth of supermassive black holes in the early universe.

\acknowledgments

This work was supported by the National Research Foundation of Korea (NRF) grant, No. 2008-0060544,
funded by the Korean government (MSIP).
The Gemini data were taken through the K-GMT Science Program (PID: KR-2015B-005) of Korea Astronomy and Space Science Institute (KASI).
Based on observations obtained at the Gemini Observatory 
acquired through the Gemini Observatory Archive and processed using the Gemini IRAF package,
which is operated by the Association of Universities for Research in Astronomy, Inc., under a cooperative agreement with the NSF on behalf of the Gemini partnership: the National Science Foundation (United States), the National Research Council (Canada), CONICYT (Chile), Ministerio de Ciencia, Tecnolog\'{i}a e Innovaci\'{o}n Productiva (Argentina), and Minist\'{e}rio da Ci\^{e}ncia, Tecnologia e Inova\c{c}\~{a}o (Brazil).
This paper includes data taken at The McDonald Observatory of The University of Texas at Austin.
Based on observations at Kitt Peak National Observatory, National Optical Astronomy Observatory 
(NOAO Prop. ID: 2012B-0537, 2013A-0506, 2013B-0534; PI: Y. Jeon), 
which is operated by the Association of Universities for Research in Astronomy (AURA) under 
cooperative agreement with the National Science Foundation.
The authors are honored to be permitted to conduct astronomical research on Iolkam Du'ag (Kitt Peak), a mountain with particular significance to the Tohono O'odham.
This paper includes data gathered with the 6.5 meter Magellan Telescopes located at Las Campanas Observatory, Chile.
M.H. acknowledges the support from Global Ph.D. Fellowship Program through the National Research Foundation of Korea (NRF) funded by the Ministry of Education (NRF-2013H1A2A1033110).
H.D.J is supported by an appointment to the NASA Postdoctoral Program at 
the Jet Propulsion
Laboratory, administered by Universities Space Research Association under contract with NASA.
D.K. acknowledges the fellowship support from the grant NRF-2015-Fostering
Core Leaders of Future Program, No. 2015-000714 funded by the Korean 
government. We thank the anonymous referee for useful comments, which improved the content of this paper.

{\it Facilities:} \facility{Mayall (RCSPL)}, \facility{NTT (EFOSC2)},    
\facility{Magellan:Baade (FIRE)}, \facility{Gemini:North (GNIRS)} 

%

\end{document}